\documentclass{gGAF2e}
\usepackage{color}

\begin{document}

\jvol{00} \jnum{00} \jyear{2020} 



\markboth{\rm Y.~LIN AND J.~NOIR}{\rm GEOPHYSICAL \&  ASTROPHYSICAL FLUID DYNAMICS}

\title{Libration-driven inertial waves and mean zonal flows in spherical shells}

{\author{Yufeng Lin ${\dag}$$^{\ast}$\thanks{$^\ast$Corresponding author. Email: linyf@sustech.edu.cn
\vspace{6pt}} and J\'er\^ome Noir${\ddag}$\\
\vspace{6pt}  ${\dag}$ Department of Earth and Space Sciences, Southern University of Science and Technology, Shenzhen 518055, China \\
 ${\ddag}$ Institute of Geophysics, ETH Zurich, Zurich 8092, Switzerland \\
\vspace{6pt}\received{??} }}

\maketitle

\begin{abstract}
Several planetary bodies in our solar system undergo a forced libration owing to gravitational interactions with their orbital companions, leading to complex fluid motions in their metallic liquid cores or subsurface oceans. In this study, we numerically investigate flows in longitudinally librating spherical shells. We focus on the Ekman number dependencies of several shear layers when the libration frequency is less than twice of the rotation frequency and the libration amplitude is small. Time-dependent flows mainly consist of inertial waves excited at the critical latitudes due to the Ekman pumping singularities, forming conical shear layers. 
In particular, previous theoretical studies have proposed different scalings for the conical shear layers spawned from the critical latitudes at the inner boundary. Our numerical results favor the velocity amplitude scaling $\mathrm{O}(\varepsilon E^{1/12})$ predicted by Le Diz\`es \& Le Bars ({\it J.~Fluid Mech.} 2017, {\bf 826}, 653) over the scaling  $\mathrm{O}(\varepsilon E^{1/6})$ initially proposed by Kerswell  ({\it J.~Fluid Mech.} 1995, {\bf 298}, 311), though the Ekman numbers in our calculations are not sufficiently small to pin down this scaling.
Non-linear interactions in the boundary layers drive a mean zonal flow with several geostrophic shears.  Our numerical results show that geostrophic shears associated with the critical latitudes at the inner and outer boundaries exhibit the same scalings, i.e. an amplitude of  $\mathrm{O}(\varepsilon^2 E^{-1/10})$ over a width of $\mathrm{O}(E^{1/5})$.  Apart from the geostrophic shear associated with the critical latitude, our numerical results show that the reflection of inertial waves can induce a geostrophic shear with an amplitude of $\mathrm{O}(\varepsilon^2 E^{-1/6})$ over a width of $\mathrm{O}(E^{1/3})$. As the amplitude of the geostrophic shears increases as reducing the Ekman number, the geostrophic shears in the mean flows may be significant in planetary cores and subsurface oceans given small Ekman numbers of these systems.
  

\end{abstract}

\section{Introduction}\label{sec:intro}
As a result of gravitational interactions with their orbital companions, synchronized (or pseudo-synchronized) bodies usually undergo a forced libration in longitude, i.e. a harmonic oscillation of the rotation rate \citep{Comstock2003}. Many librating bodies in our solar system contain large volume of fluids either in the form of a metallic liquid core (e.g. Mercury, Io, Ganymede, and the Moon) or a subsurface ocean (e.g. Europa, Titan, Callisto, Ganymede and Enceladus) \citep[see][ and references therein]{Noir2009}. The librating solid shells can excite complex and even turbulent flows in the fluid layers through viscous and topographic couplings \citep[see the review by][and references therein]{LeBars2014}. Understanding fluid responses in librating bodies is crucial to study dynamics and evolution of these planetary bodies. For instance, energy dissipation resulting from libration-driven turbulence may provide heat source to maintain subsurface oceans of icy satellites \citep{Wilson2018}. Furthermore, libration has been proposed as a possible mechanism to drive planetary dynamos \citep{LeBars2011Nature}. 

Motivated by the aforementioned applications in the dynamics of planetary interiors, a great deal of attention has been given to flows driven by libration over the last decade. When the libration frequency is less than the twice of the rotation frequency, smooth inertial modes can be exited in a container where regular inertial modes exist, such as in a sphere \citep{Aldridge1969,Zhang2013} or  ellipsoids \citep{  Vantieghem2015}. {  In a spherical shell, however, regular inertial modes generally do not exist and inertial wave beams are generated at the critical latitudes due to the singularities of the oscillatory viscous boundary layers \citep{Kerswell1995, Rieutord1997}. The thin wave beams propagate in the bulk of fluid along the characteristic surfaces, forming the conical shear layers  \citep{Tilgner1999,Calkins2010,Koch2013,Hoff2016}. In certain frequency bands, the inertial wave beams converge to wave attractors after multiple reflections \citep{Rieutord2001,Ogilvie2009,Koch2013}. When the frequency is $2 \sin (p \pi/q)$,  where $p$ and $q$ are integers, the wave beam is periodic and there is no attractor \citep{Rieutord2001, Rieutord2018}. At these particular frequencies, the conical shear layers spawned form the critical latitudes lead to a simple closed trajectory, the so-called periodic orbit \citep{Rieutord2001}. In this study, we consider only cases of periodic orbits and set the libration frequency $\omega_L=1.0$ and $\omega_L=\sqrt 2$ to investigate the conical shear layers spawned from the critical latitudes in spherical shells. } 

As a consequence of non-linear interactions of time-dependent flows, a steady mean flow in azimuthal direction (i.e. zonal flow) can be generated. Libration-driven mean zonal flows have been observed both experimentally \citep{Noir2010,Sauret2010,Noir2012,Seelig2015} and numerically \citep{Tilgner2007b,Calkins2010,Sauret2010}. In the absence of inertial waves, non-linear interactions in viscous boundary layers drive a zonal flow whose amplitude is proportional to the square of the libration amplitude and is independent of the Ekman number \citep{Busse2010,Sauret2012}. However, if the libration frequency is less than twice of the rotation frequency, the excitation of inertial waves complicates the structure of mean zonal flows, but detailed interactions between inertial waves and mean flows remain to be elucidated. \cite{LeDizes2015a} obtained an analytical solution of the zonal flow generated by the non-linear interactions of inertial waves for a librating disk in an unbounded fluid domain. In an enclosed container, however, it is very challenging to tackle the problem analytically owing to complicated reflections of inertial waves on the boundary \citep{Sauret2012}. 

{When the libration amplitude is sufficiently large, libration-driven flows become unstable and eventually lead to turbulence \citep{Noir2009}. Several instability mechanisms have been identified and studied both experimentally and numerically \citep{Noir2009,Chan2011Lon,Cebron2012,Sauret2013,Grannan2014,Favier2015,Lemasquerier2017}. For libration-driven instabilities, we refer the reader to the review by \cite{LeBars2014}. }

{In this study, we focus on the Ekman number dependencies of time-dependent conical shear layers and geostrophic shears in the mean zonal flows when the libration amplitude is small. We perform a set of axisymmetric numerical simulations in librating spherical shells at two libration frequencies $\omega_L=1.0$ and $\omega_L=\sqrt {2}$, at which the conical shear layers form simple ray paths after few reflections. We investigate scalings of the conical shear layers and compare our numerical results with previous theoretical predictions \citep{ Kerswell1995, LeDizes2017}.}  Several geostrophic shears are observed in the mean zonal flows. Apart from a dominant geostrophic shear associated with the critical latitudes where inertial waves are initially launched, our numerical results reveal that reflections of inertial waves on viscous {boundaries} induce geostrophic shears as well.  The amplitude of all of these geostrophic shears increases as the Ekman number is decreased. This may have significant implications for the dynamics of liquid cores and subsurface oceans of planetary bodies as Ekman numbers are extremely small at planetary settings.      

The remaining part of this paper is organized as follows. Section \ref{sec3:model} introduces the setup of the problem and the numerical scheme. Section \ref{sec3:result} presents our numerical results. We close the paper with a summary and discussion in section \ref{sec3:diss}.

\section{Numerical model}\label{sec3:model}
\begin{figure}
\begin{center}
\includegraphics[width=0.45\textwidth]{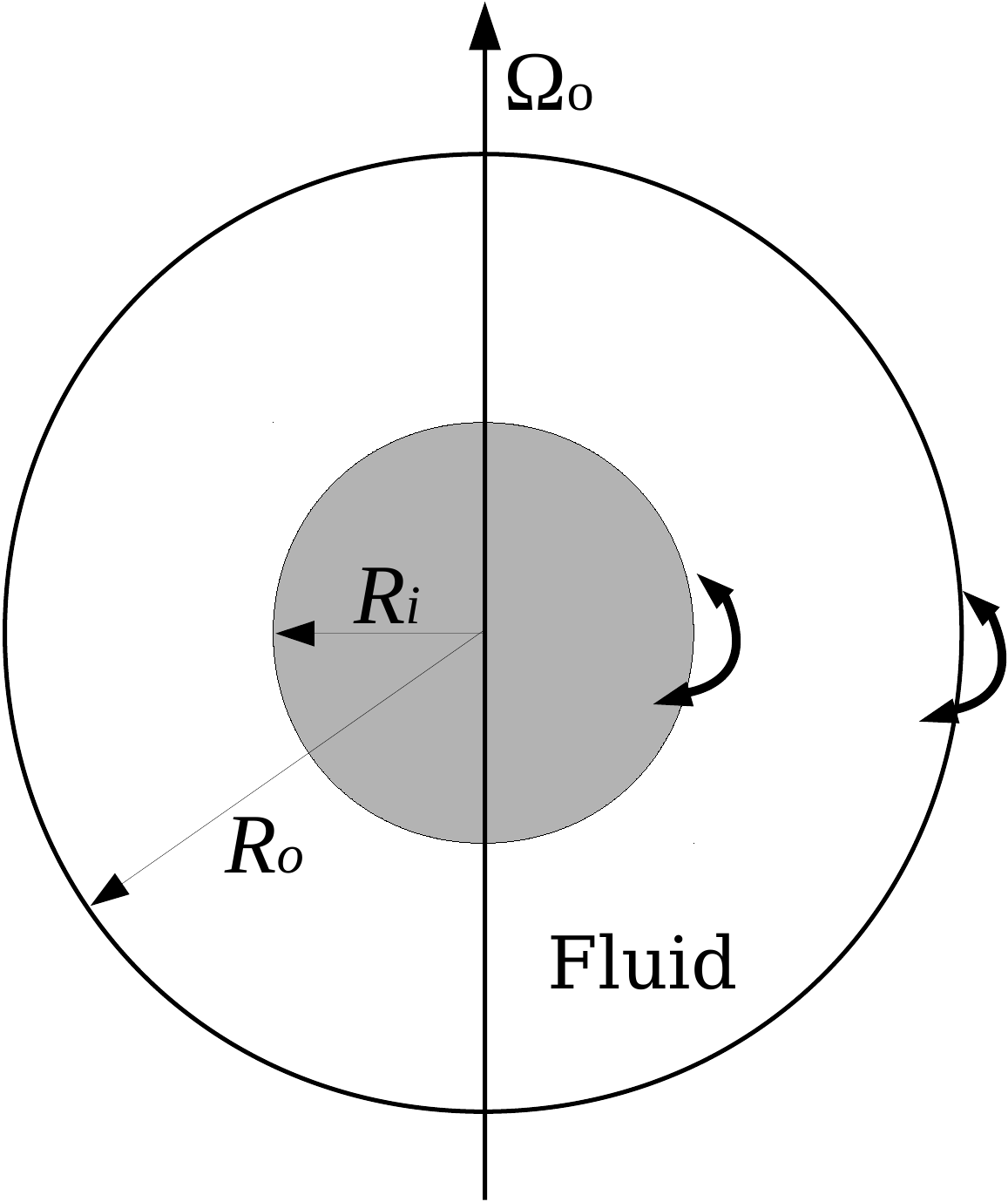}
\end{center}
\caption{Sketch of the problem.}
\label{fig1}
\end{figure}

We consider a spherical shell of outer radius $R_o$ and inner radius $R_i$, enclosed with an incompressible and homogeneous fluid with density $\rho$ and kinematic viscosity $\nu$ ({ figure \ref{fig1}}). The radius ratio of the spherical shell is given as $r_i=R_i/R_o$. The outer shell and the inner core rotate at a mean angular velocity $\varOmega$, with additional oscillations such that
\begin{align}
\varOmega_{o}(t)\,=\,&\,\varOmega+\Delta\phi_{o}\varOmega_L\sin(\varOmega_Lt)\,,\label{eq3:omega}\\
\varOmega_{i}(t)\,=\,&\,\varOmega+\Delta\phi_{i}\varOmega_L\sin(\varOmega_Lt)\,,
\end{align}
where $\varOmega_L$ is angular libration frequency, $\Delta\phi_{o}$ and  $\Delta\phi_{i}$ {are} the libration amplitude of the outer shell and inner core respectively. Using $\varOmega^{-1}$ as a time scale and $R_o$ as a length scale, the Navier-Stokes equation in the frame rotating with the mean angular velocity is given by 
\begin{equation}\label{eq3:NS}
\frac{\upartial \bm u}{\upartial t}+\bm{u \cdot \nabla u}+2\bm{\hat z \times u}\,=\,-\,\bm \nabla p+E\nabla^2\bm u\,,
\end{equation}
where $\bm u$ is the velocity, $\bm{\hat z}$ is the unit vector parallel to the rotation axis, $p$ is modified pressure including the centrifugal term and $E=\nu/(\varOmega R_o^2)$ is the Ekman number. The mass conservation is written as 
\begin{equation}
\bm{ \nabla \cdot u}\,=\,0\,.
\end{equation}

In spherical coordinates ($r,\theta,\phi$), the no-slip boundary condition on both surfaces gives 
\begin{equation}\label{eq3:BC1}
u_r\,=\,u_\theta\,=\,0\,, \hskip 10mm  \text{at} \hskip 5mm r\,=\,r_i \hskip 3mm\text{and}\hskip 3mm r\,=\,1\,.
\end{equation} 
Libration is imposed through the boundary conditions of $u_\phi$. When the outer shell is librating, 
\begin{equation}
u_\phi|_{r=1}\,=\,\varepsilon_{o} \cos \theta \sin(\omega_L t)\,,
\end{equation}
and $u_\phi|_{r=1}=0$ otherwise. 
When the inner core is librating, 
\begin{equation}
u_\phi|_{r=r_i}\,=\,r_i\varepsilon_{i} \cos \theta \sin(\omega_L t)\,,
\end{equation}
and $u_\phi|_{r=r_i}=0$ otherwise. 
Here $\omega_L=\varOmega_L/\varOmega$ is the  dimensionless libration frequency, $\varepsilon_{o}=\Delta \phi_{o} \omega_L$  and $\varepsilon_{i}=\Delta \phi_{i} \omega_L$ are dimensionless libration  amplitudes of the outer shell and the inner core, respectively. 

As we are primarily interested in the mean zonal flow generated by the non-linear interactions of inertial waves, we restrict our numerical simulations to weak libration amplitude regime such that there is no instabilities arising in the system. This allows us to carry out axisymmetric simulations and reach relatively low Ekman number ($E=10^{-7}$). We use a numerical code developed by \cite{Calkins2010} for axisymmetric simulations of libration-driven flows in a spherical geometry. The Navier-Stokes equations together with the boundary conditions are solved using a vorticity-stream function formulation in the meridional plane, which can eliminate the pressure and guarantee the divergence free of the velocity.
The azimuthal velocity, the azimuthal vorticity and the stream function are expanded as Legendre polynomials in latitude and discretized using a second-order finite difference in radius. The finite differences in radius choose Gauss-Lobatto points to increase the resolution next to the boundaries.
For the time evolution, a second-order Adams-Bashforth backward differentiation scheme is used. The numerical code has been benchmarked and more details about the numerical scheme can be found in \cite{Calkins2010}.  
\section{Results} \label{sec3:result}
We carry out three groups of simulations listed in table. \ref{tab1}. In each group, all parameters are fixed except for varying the Ekman number. In Group~1, the inner core ($r_i=0.1$) is uniformly rotating while the outer shell is librating at the frequency $\omega_L=1.0$ with the amplitude $\varepsilon=1.0\times 10^{-2}$. In Group~2, the outer shell is uniformly rotating while the inner core ($r_i=0.35$) is librating at the frequency $\omega_L=1.4142$ with the amplitude $\varepsilon=1.4142\times 10^{-2}$. In Group~3, the inner core  ($r_i=0.35$) is uniformly rotating while the outer shell is librating at the frequency $\omega_L=1.4142$ with the amplitude $\varepsilon=1.4142\times 10^{-2}$.  { These particular frequencies correspond to simple periodic orbits, which lead to simple beam structures of the conical shear layers.} The libration amplitude is small to ensure that no instabilities are triggered.

\begin{table}
\caption{Parameters of the numerical model setup.}
\label{tab1}
\begin{center}

\begin{tabular}{lcccccc}
\hline\noalign{\smallskip}
Model & $\omega_L$ & $\varepsilon_{o}$ & $\varepsilon_{i}$ & $r_i$  & $E$ \\
\hline\noalign{\smallskip}
Group~1 & 1.0 & $1.0 \times 10^{-2}$ & 0 & 0.1 & $10^{-5}-10^{-7}$ \\
Group~2 & 1.4142 & 0 &$1.4142 \times 10^{-2}$ & 0.35 & $10^{-5}-10^{-7}$ \\
Group~3 & 1.4142 & $1.4142 \times 10^{-2}$ & 0 & 0.35 & $10^{-5}-10^{-7}$ \\
\noalign{\smallskip}\hline
\end{tabular}
\end{center}
\end{table}

\subsection{Time-dependent shear layers}
In the stable regime with $\omega_L<2$, the time-dependent flow in a librating spherical shell is mainly characterized by inertial waves launched from the critical latitudes, where the Ekman boundary layer is singular \citep{Stewartson1957}. Inertial waves propagate into the bulk of the fluid and form internal shear layers along the characteristics, which are cones with a fixed open angle with respect to the rotation axis \citep{Kerswell1995}. The angle is solely determined by the libration frequency and is given as $\theta_c=\arcsin \left(\omega_L/2 \right)$ \citep{Greenspan1968}. The conical shear layers can be spawned from either the critical latitudes on the outer boundary (concave surface), namely the Core-Mantle Boundary (CMB) or from the critical latitudes on the inner boundary (convex surface), namely the Inner Core Boundary (ICB). 

\subsubsection{Oscillating conical shear layers spawned from the critical latitude at the CMB.}
\begin{figure}
\begin{center}
\includegraphics[width=0.45 \textwidth,clip,trim={2cm 0 0 1cm} ]{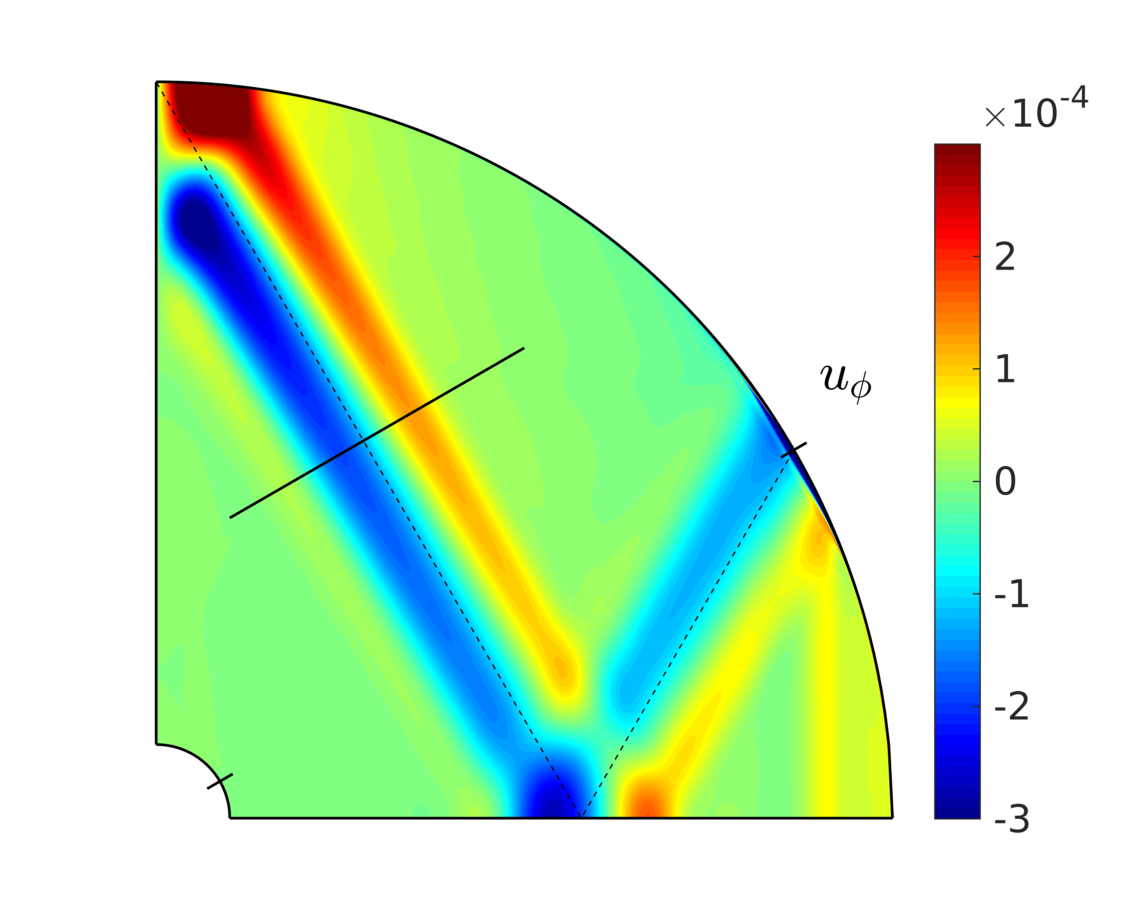}
\includegraphics[width=0.5 \textwidth ]{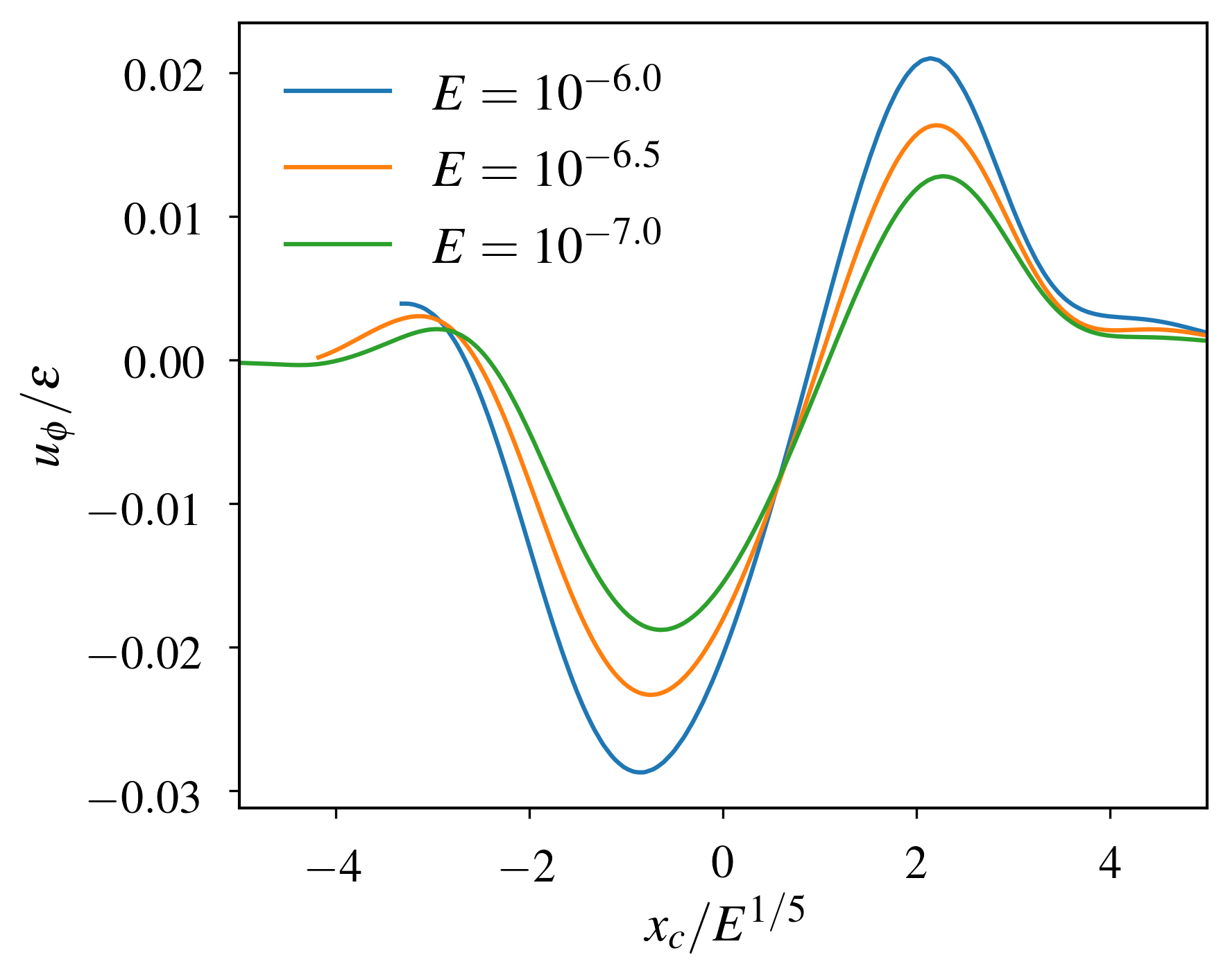} \\
\hspace*{3cm} (a) \hfill (b) \hspace*{4cm}
\caption{(a) Azimuthal velocities $u_\phi$  in the meridional plane at $E=10^{-7}$ in Group~1. Tick marks show the critical latitude and dashed lines show the characteristic surfaces. The color scale in the boundary layers is saturated. (b) Azimuthal velocity $u_\phi$ profiles along the solid line in (a) at different Ekman numbers. A local coordinate $x_c$ along the solid line is used and $x_c=0$ is set at the intersection of the solid line and the characteristic surface. $u_\phi$ and $x_c$ are divided by the libration amplitude $\epsilon$ and $E^{1/5}$ respectively (Colour online).}
\label{fig2}
\end{center}
\end{figure}

\begin{figure}
\begin{center}
\includegraphics[width=0.95 \textwidth]{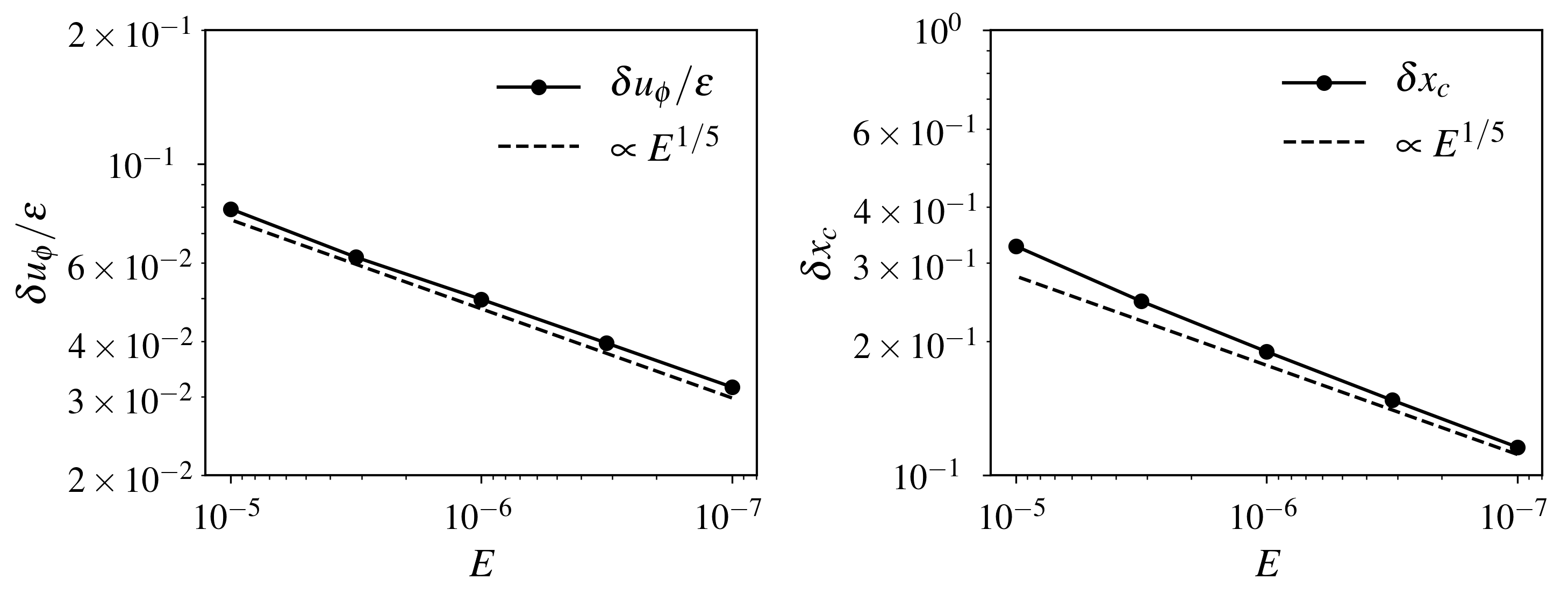} \\
\hspace*{4cm}(a)\hfill (b) \hspace*{3cm}
\caption{{(a) Peak-to-peak velocity $\delta u_\phi$ divided by the libration amplitude $\varepsilon$, and (b) the distances between the peaks $\delta x_c$ as a function of the Ekman number for the conical shear layers in figure \ref{fig2} (Group~1). }}
\label{fig3}
\end{center}
\end{figure}

First we consider a case in Group~1, where the outer shell is librating at $\omega_L=1.0$ while the inner core is uniformly rotating. We use a small inner core ($r_i=0.1$) in this group to avoid that the conical shear layers spawned from the outer critical latitude touch the inner core. Figure \ref{fig2}(a) shows azimuthal velocities in a meridional plane when the outer shell is at the maximum retrograde position for $E=10^{-7}$. Owing to the axi-symmetry and equatorial symmetry, we plot only one quarter of the meridional plane here.  Conical shear layers spawned from the critical latitude at the outer boundary propagate along the characteristics and reach the poles, forming a very simple ray path. Note that the reflection on the equatorial plane is due to the mirror symmetry around the equator rather than a physical reflection.

{ Figure \ref{fig2}(b) shows the azimuthal velocity $u_\phi$, divided by the libration amplitude $\varepsilon$, along a line perpendicular to the conical shear layers at three different Ekman numbers in the range of $[10^{-6}, 10^{-7}]$. We defined a local coordinate $x_c$ along the solid line and set $x_c=0$ at the intersection of the solid line and the characteristic surface (represented by the oblique dashed line in figure \ref{fig2}(a)). The local coordinates $x_c$ is rescaled by $E^{1/5}$ in figure \ref{fig2}(b). We can see that the amplitudes of the shears decrease as reducing the Ekman number $E$. 

In order to check the scaling of the width and the velocity amplitude of  the shear layers, 
we pick up the peak-to-peak velocities $\delta u_\phi$ (the difference between the maximum and minimum) along the solid line crossing the shear, and the distances $\delta x_c$ between the peaks. Figure \ref{fig3} shows $\delta u_\phi/\varepsilon$ and $\delta x_c$ as a function of the Ekman number in the range of $[10^{-5}, 10^{-7}]$. We can see that numerical results are in very good agreement with the theoretical scalings that the conical shear layers spawned from the critical latitudes at the CMB have a length scale $\mathrm{O}(E^{1/5})$ carrying velocities $\mathrm{O}(\varepsilon E^{1/5})$ \citep{Roberts1963,Kerswell1995,Noir2001,Kida2011}.} \footnote{\cite{Kerswell1995} proposed the length scale $\mathrm{O}(E^{1/5})$ but a velocity amplitude $\mathrm{O}(\varepsilon E^{3/10})$.}

\subsubsection{Oscillating conical shear layers spawned from the critical latitude at the ICB}
\begin{figure}
\begin{center}
{\includegraphics[width=0.45 \textwidth, clip,trim={2cm 0 0 1cm}]{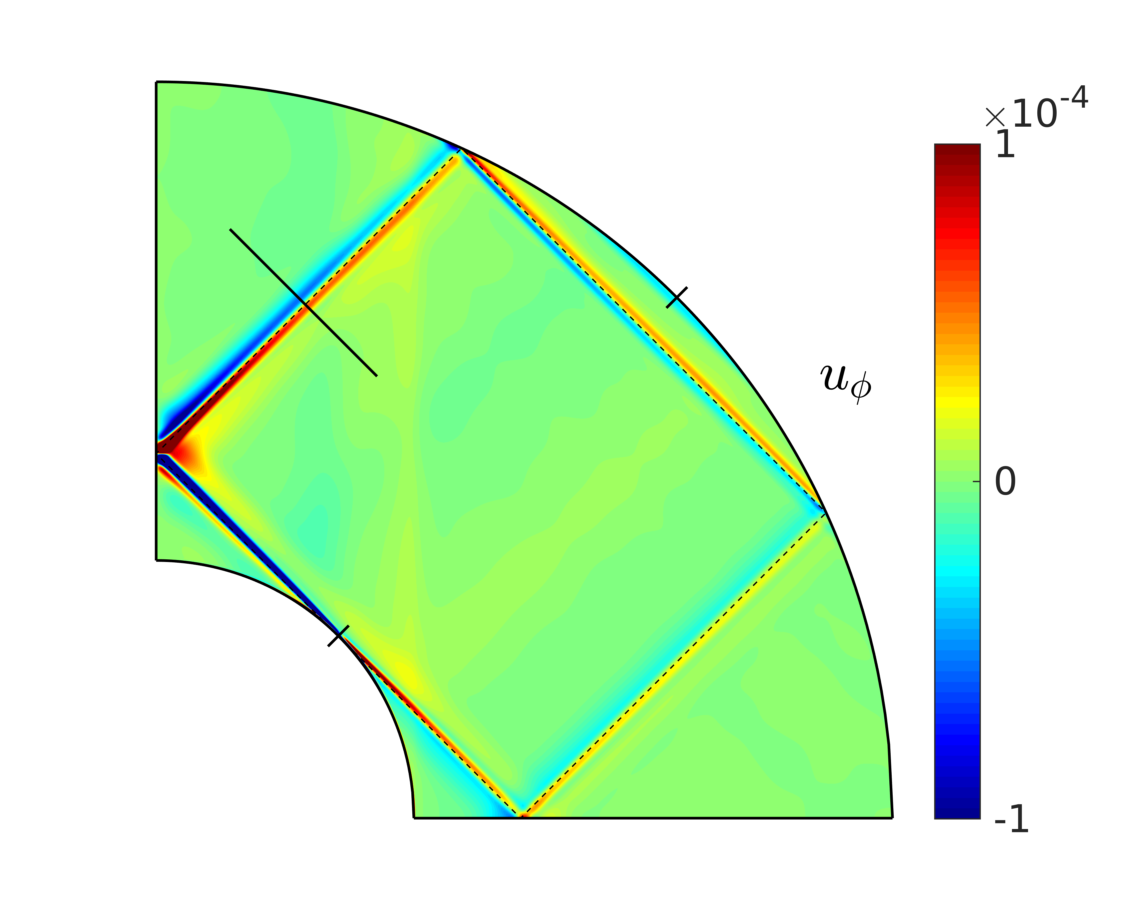}}
\hfill
{\includegraphics[width=0.50 \textwidth]{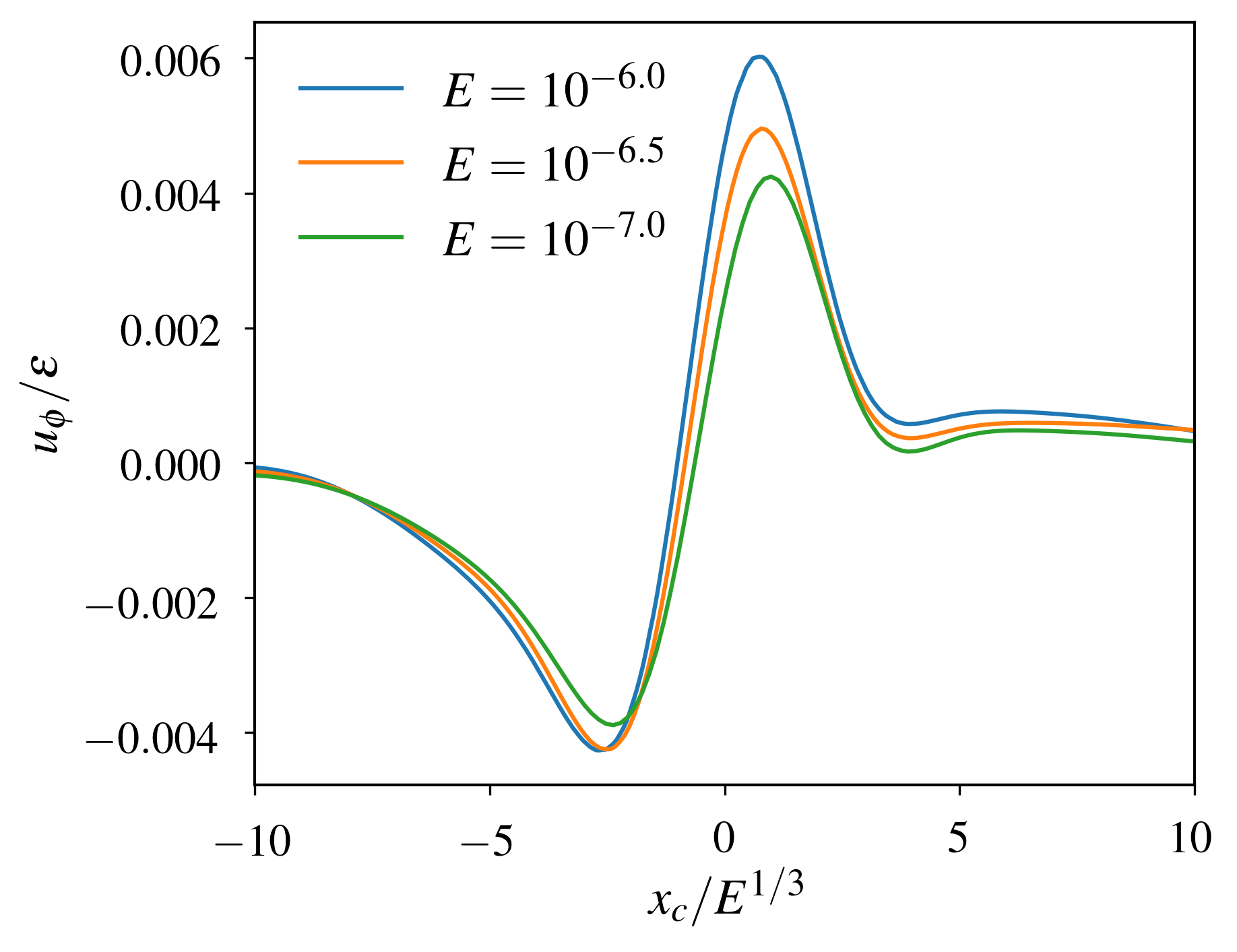}}  \\
\hspace*{3cm} (a) \hfill (b) \hspace*{4cm}
\caption{(a) Azimuthal velocities $u_\phi$  in the meridional plane at $E=10^{-7}$ in Group~2. Tick marks show the critical latitude and dashed lines show the characteristic surfaces. The color scale in the boundary layers is saturated. (b) Azimuthal velocity $u_\phi$ profiles along the solid line in (a) at different Ekman numbers. A local coordinate $x_c$ along the cross line is used and $x_c=0$ is set at the intersection (close to the north pole) of the solid line and the characteristic surface. $u_\phi$ and $x_c$ are divided by $\varepsilon$ and $E^{1/3}$ respectively. Only the section close to the north pole is shown here (Colour online).}
\label{fig4}
\end{center}
\end{figure}

We now consider a case in Group~2 where the inner core librates at a frequency $\omega_L=\sqrt{2}$ and the outer shell rotates uniformly. This model is designed to study the conical shear layers spawned from the critical latitudes on the inner boundary. The particular libration frequency $\omega_L=\sqrt{2}$ is chosen { so that the inertial waves beam form a periodic trajectory but not an attractor \citep{Rieutord2001}.} The flow pattern is illustrated in figure \ref{fig4}(a), showing azimuthal velocities in a meridional plane when the inner core is at the maximum retrograde position for $E=10^{-7}$. In contrast with the previous case, the conical shear layers are spawned tangentially from the inner critical latitude ($\theta_c=\pi/4$), forming a simple { periodic orbit with a rectangular ray path comprising 2 reflections at the CMB in each hemisphere}

\begin{figure}
\begin{center}
\includegraphics[width=0.95 \textwidth]{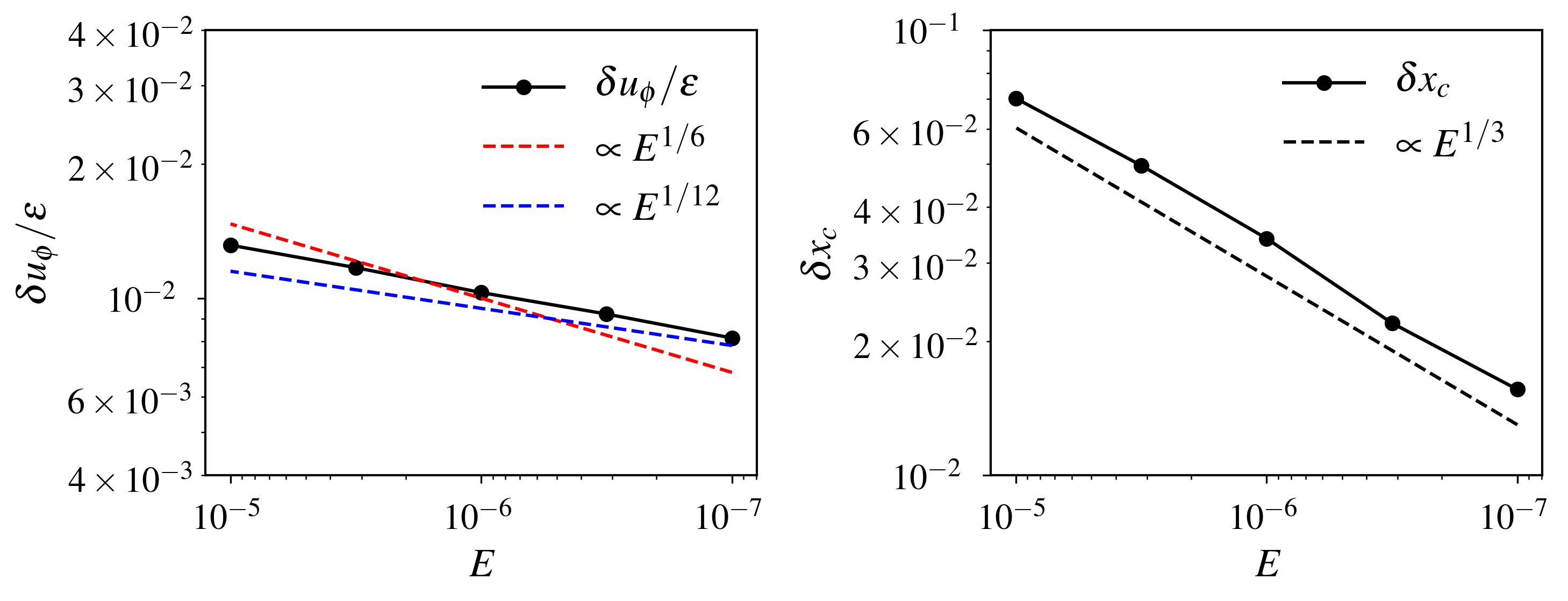}\\
\hspace*{4cm}(a)\hfill (b) \hspace*{3cm}
\caption{{(a) Peak-to-peak velocity $\delta u_\phi$ divided by the libration amplitude $\varepsilon$, and (b) the distances between the peaks $\delta x_c$as a function of the Ekman number for the conical shear layers in figure \ref{fig4} (Group~2) (Colour online). }}
\label{fig5}
\end{center}
\end{figure}

In figure \ref{fig4}(b) we plot the azimuthal velocity $u_\phi$ crossing the shear layer at three different Ekman numbers in the range of $[10^{-6}, 10^{-7}]$. As previously, we introduce a local coordinate $x_c$ perpendicular to the characteristic line. $u_\phi$ and $x_c$ are divided by the libration amplitude $\varepsilon$ and $E^{1/3}$ (i.e. natural length scale in rotating fluids)  respectively. \cite{Kerswell1995} proposed that the velocity amplitude of the conical shear layer tangential to the critical latitude should scale as $\varepsilon E^{1/6}$ over a width of $E^{1/3}$.  Recently, \cite{LeDizes2017} argued that the velocity amplitude of the conical shear layers from a convex boundary should scale as $\varepsilon E^{1/12}$ in an unbounded fluid domain. 
In order to examine the Ekman number dependencies in detail, we show the peak-to-peak velocity $\delta u_\phi$ crossing the shear, divided by the libration amplitude $\varepsilon$,  and the distance  $\delta x_c$ between the peaks as a function of the Ekman number in the range of $[10^{-5}, 10^{-7}]$ in figure \ref{fig5}. We can see that the width of the conical shear layers  spawned from the critical latitude at the ICB scales as $E^{1/3}$ as expected \citep{Kerswell1995}. Regarding the velocity amplitude, it turns out that the scaling of $\varepsilon E^{1/12}$ fits our numerical results better than $\varepsilon E^{1/6}$, { yet our numerical calculations may not be in a fully asymptotic regime even at $E=10^{-7}$}.

\begin{figure}
\begin{center}
\includegraphics[width=0.45 \textwidth,clip,trim={2cm 0 0 1cm}]{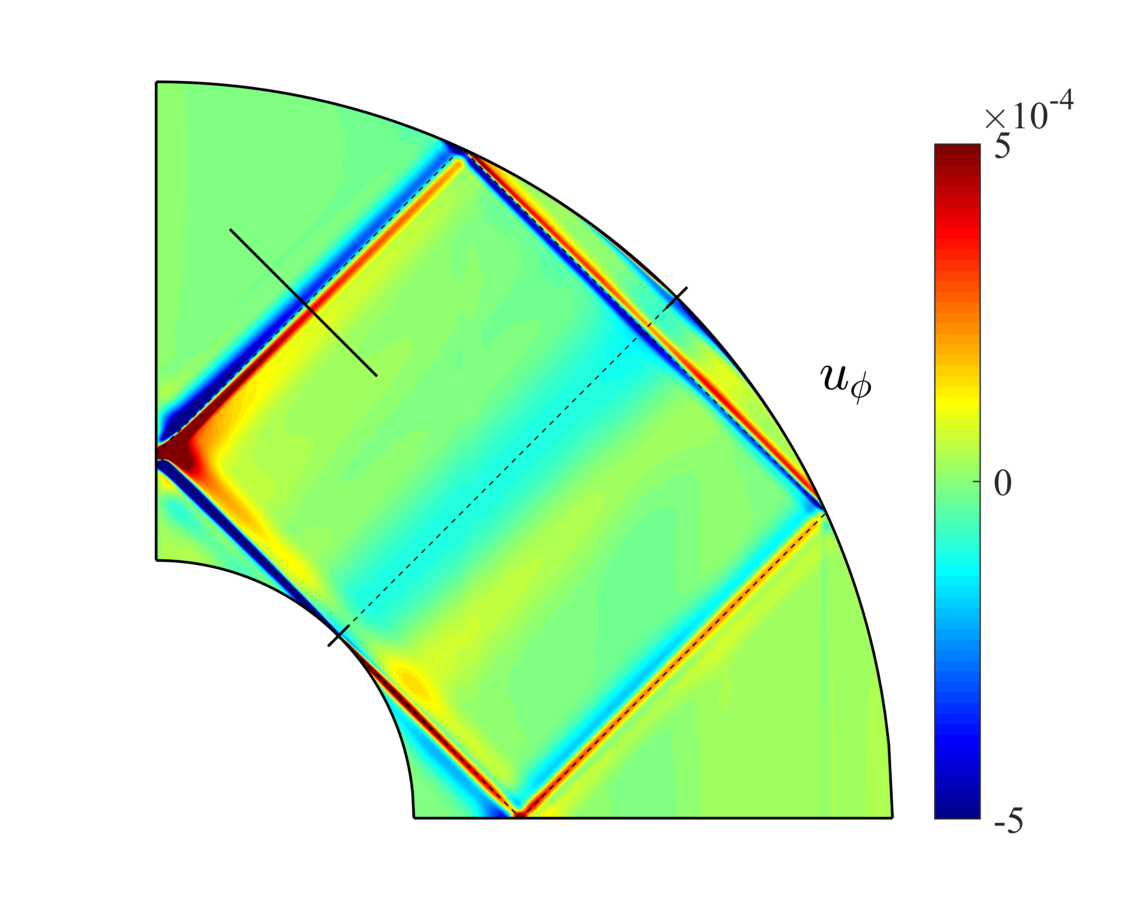}
\includegraphics[width=0.54 \textwidth]{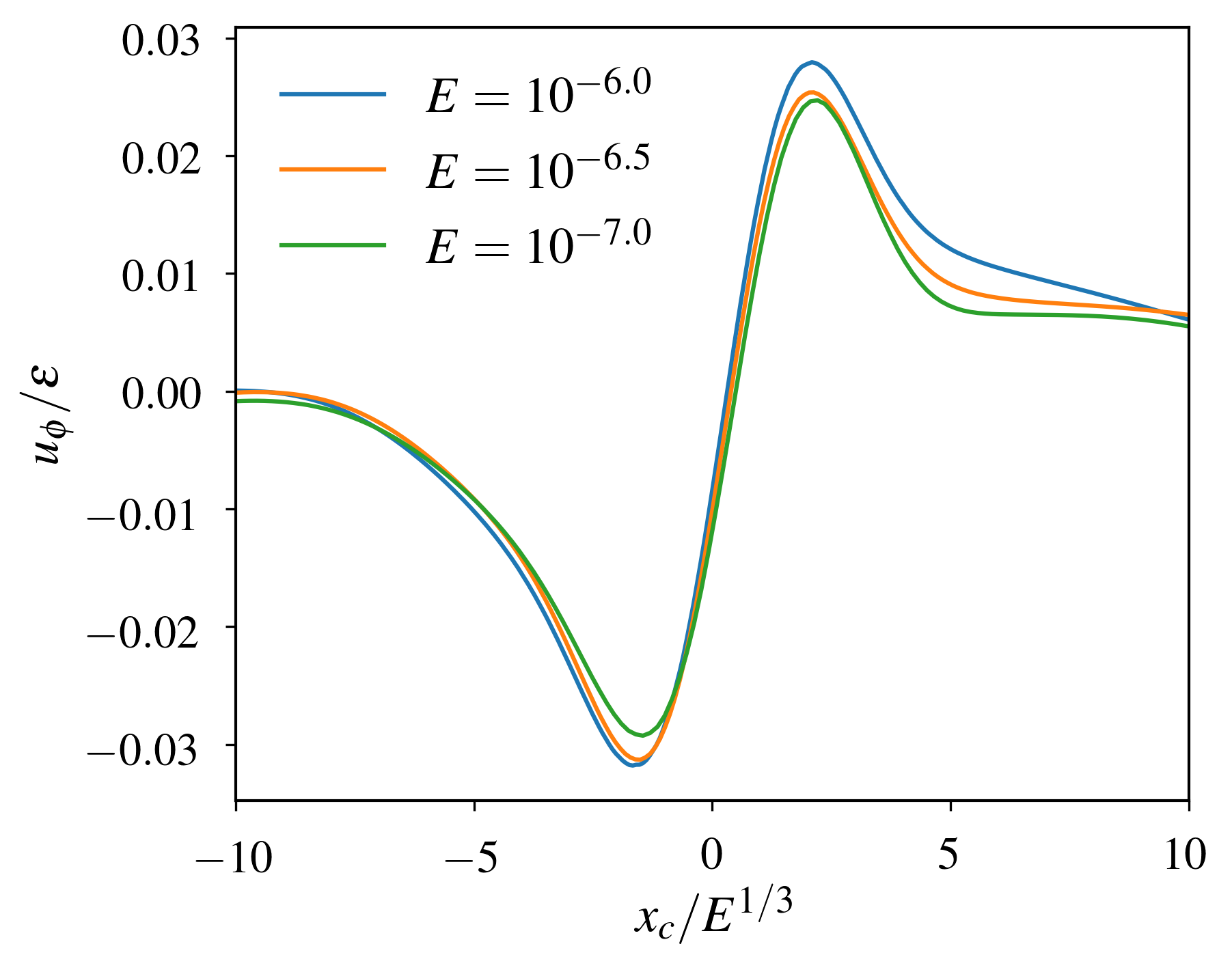} \\
\hspace*{3cm} (a) \hfill (b) \hspace*{4cm}
\caption{(a) Azimuthal velocities $u_\phi$  in the meridional plane at $E=10^{-7}$ in Group~3. Tick marks show the critical latitude and dashed lines show the characteristic surfaces. The color scale in the boundary layers is saturated. (b) Azimuthal velocity $u_\phi$ profiles along the solid line in (a) at different Ekman numbers. A local coordinate $x_c$ along the cross line is used and $x_c=0$ is set at the intersection (close to the north pole) of the solid line and the characteristic surface. {$u_\phi$ and $x_c$ are divided by $\varepsilon$ and $E^{1/3}$ respectively.} Only the section close to the north pole is shown here (Colour online). }
\label{fig6}
\end{center}
\end{figure}

We also consider the case where the inner core is uniformly rotating while the outer shell is librating at $\omega_L=\sqrt{2}$ (Group~3). Figure \ref{fig6}(a) shows azimuthal velocities in a meridional plane at $E=10^{-7}$ in this case. As we can see, the flow is dominated by the conical shear layers spawn from the inner critical latitudes despite the uniformly rotating inner core. In fact, the conical shear layer spawned from the outer critical latitude directly hits the inner critical latitude in this case, leading to the excitation of conical shear layers tangential to the inner boundary as well.

\begin{figure}
\begin{center}
\includegraphics[width=0.9 \textwidth]{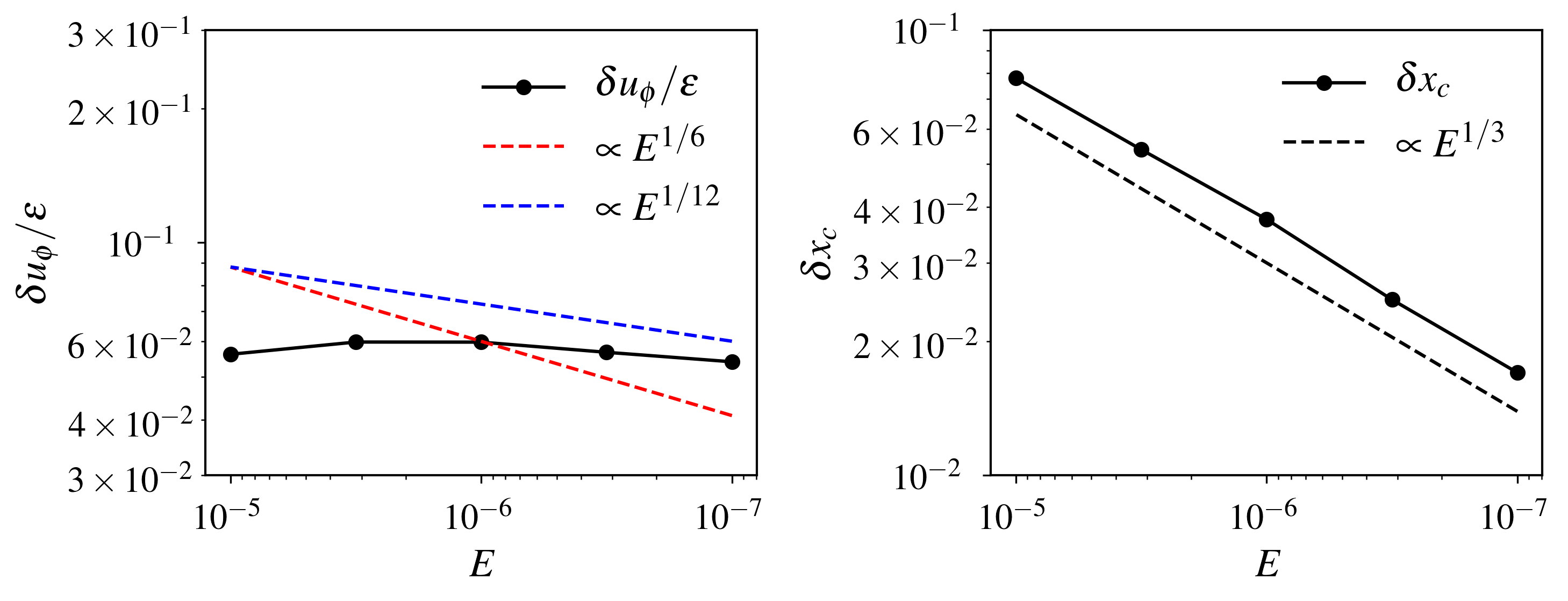}
\caption{(a) {Peak-to-peak velocity $\delta u_\phi$ divided by the libration amplitude $\varepsilon$, and (b) the distances between the peaks $\delta x_c$as a function of the Ekman number for the conical shear layers in figure \ref{fig6} (Group~3). (Colour online)}}
\label{fig7}
\end{center}
\end{figure}

We present the azimuthal velocity profile crossing the shear layer in figure \ref{fig6}(b) and the peak-to-peak scalings of the conical shear layers in figure \ref{fig7}. It is clear that the width of the conical shear layers spawned from the ICB scales as $E^{1/3}$ as previously. However, the velocity amplitude does not show asymptotic behavior yet in the range of Ekman numbers we used. Note that the conical shear layers from the ICB and from the CMB have overlap at large Ekman numbers in this case. Nevertheless, our numerical results at lower Ekman numbers are in better agreement with the scaling  $\varepsilon E^{1/12}$ rather than $\varepsilon E^{1/6}$. 

\subsection{Mean zonal flows}
\begin{figure}
\includegraphics[width=0.33\textwidth]{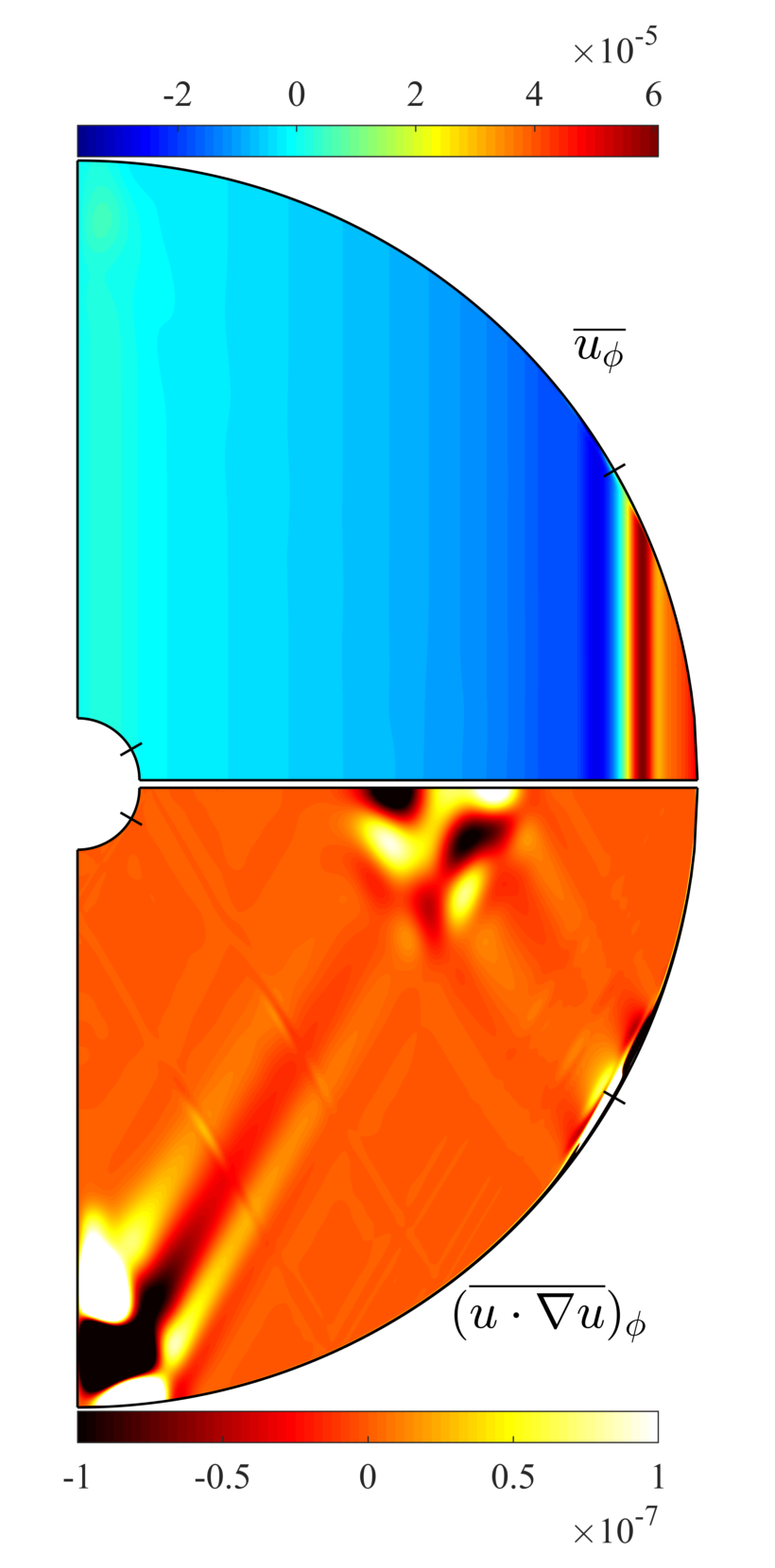}
\includegraphics[width=0.33\textwidth]{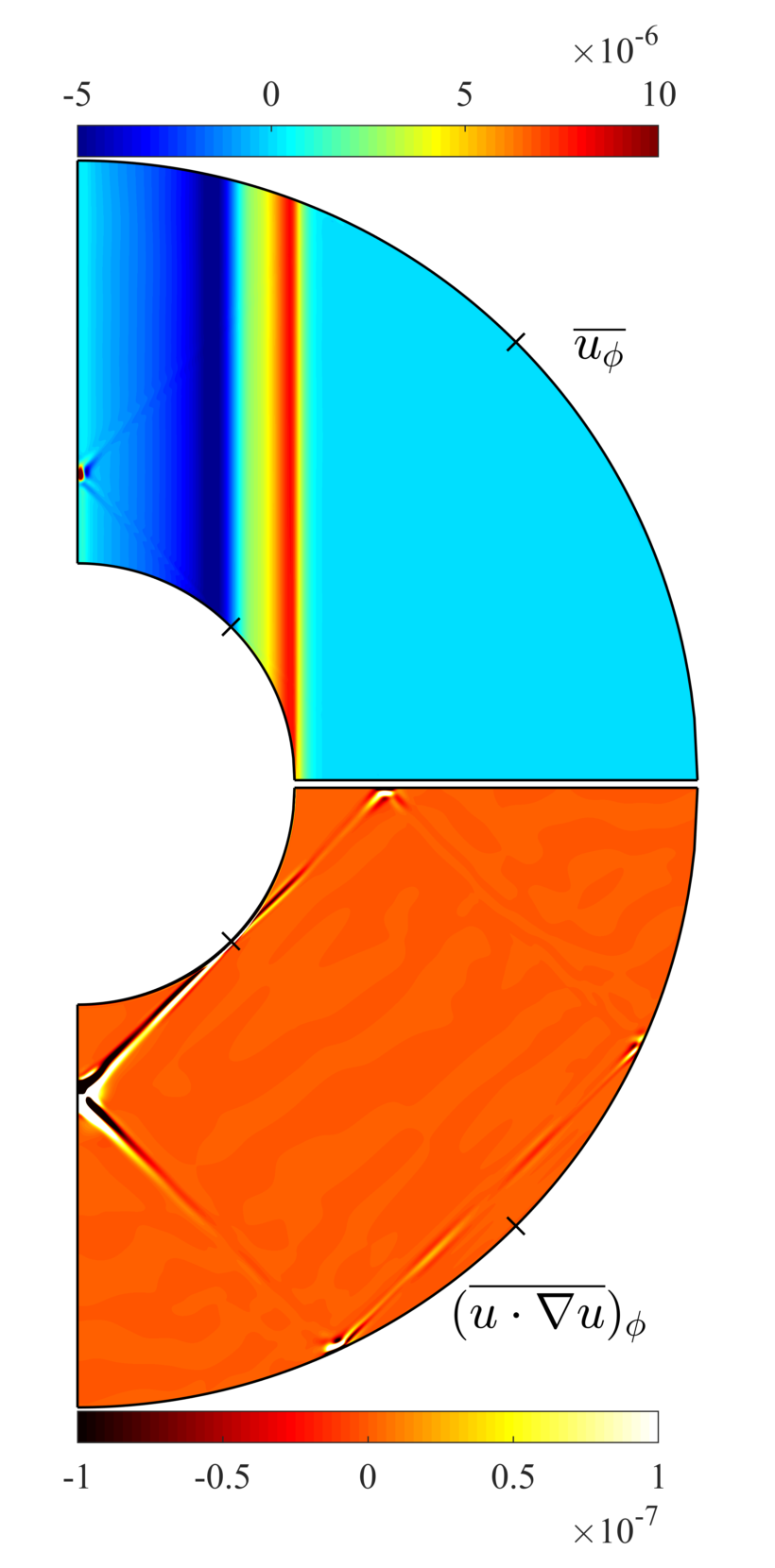}
\includegraphics[width=0.33\textwidth]{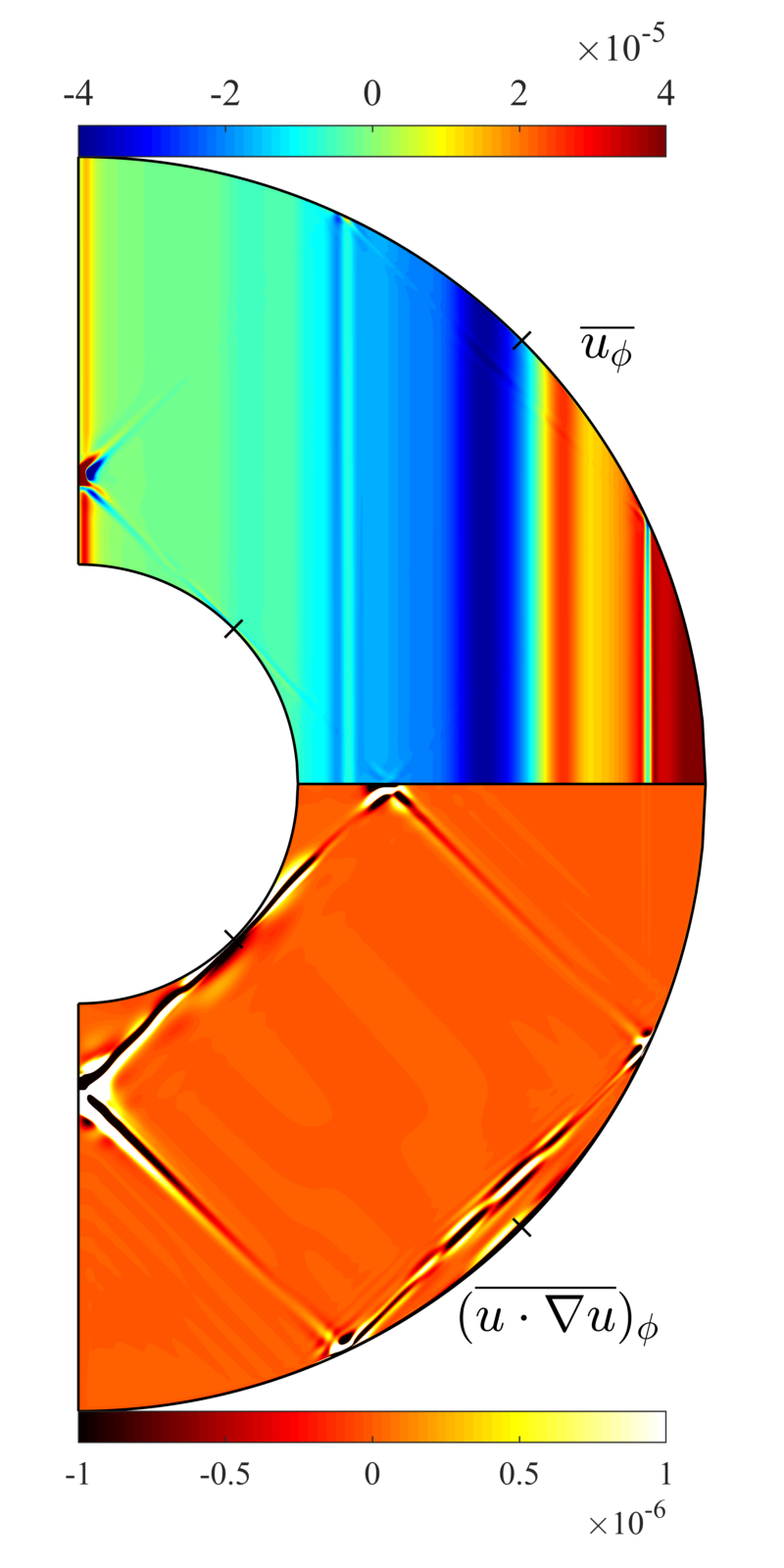} \\
\hspace*{2.5cm}(a)\hfill(b)\hfill (c) \hspace*{2.5cm} 
\caption{Time-averaged azimuthal velocities $\overline{u_\phi} $ (upper parts) and  azimuthal component of the time-averaged non-linear term $(\overline{\bm{u\cdot \nabla u}})_\phi$ (bottom parts) in the meridional plane.  {The color scale in the bottom parts is saturated in the boundary layer in order to highlight the non-linear interactions in the bulk of fluid.} Tick marks show the location of the critical latitude; $E=10^{-7}$ for all plots. (a) Group~1; (b) Group~2; (c) Group~3. (Colour online).}
\label{fig8}
\end{figure}\
The non-linear interactions of a time dependent flow $\bm u\sim \mathrm{e}^{\mathrm i \omega_L t}$ can potentially drive a steady mean flow owing to the non-linear term $\bm {u\cdot \nabla u}$ in the momentum equation \citep{Tilgner2007b}.  We calculate the mean flow $\overline{\bm u}$ by averaging the velocities over 10 libration periods after the time dependent flows are saturated. Figure \ref{fig8} shows the azimuthal components of the time-averaged velocity  $\overline{u_\phi}$ (upper parts) in the meridional plane at $E=10^{-7}$ in the three configurations discussed in the previous section. In all cases, we can see that the mean zonal flows are geostrophic, i.e. invariant along the rotation axis. 

In the case of Group~1 where the outer shell is librating at $\omega_L=1.0$, the zonal flow is characterized by a single shear layer along the cylinder associated with the critical latitude {at} the CMB (figure \ref{fig8}(a)). Such geostrophic shear was first noticed by \cite{Busse1968} when studying the steady flow in a precessing spheroid, and has been observed both in laboratory experiments \citep{Noir2001b,Morize2010} and numerical calculations \citep{Noir2001}. 

In the case of Group~2 where only the inner core is librating, we can see a geostrophic shear at the critical latitude at the ICB and another one on the tangent cylinder. Since the outer shell is not librating, the non-linear interaction in the outer boundary layer is negligible and thus there is no significant mean zonal flow outside the tangential cylinder (figure \ref{fig8}(b)). 

In the case of Group~3 where the inner core is large, the outer shell is librating at $\omega_L=\sqrt{2}$ such that the conical shear layer spawned from the critical latitude at the CMB reflects at the ICB.  In this case apart from the geostrophic shear at the outer critical latitude, there exists some additional geostrophic shears (figure \ref{fig8}(c)).
  
To illustrate the particular role played by the boundary layers in the generation of the geostrophic shears, we plot the azimuthal component of the time-averaged non-linear term $(\overline{\bm{u\cdot \nabla u}})_\phi$ in the southern hemisphere in figure \ref{fig8}. The color scale is saturated in the boundary layer in order to highlight the non-linear interactions in the bulk of fluid, which mainly take place within the conical shear layers as we can see.  We observe no systematic correlation between the time-averaged non-linear term in the bulk and the mean flow, suggesting that the mean zonal flow is mainly generated in the viscous boundary layers.    
Indeed, \cite{Greenspan1969} had shown that the non-linear interactions of inertial waves in the inviscid region have no contribution to the mean zonal flows, while a combination of viscous and non-linear effects in the viscous boundary layers can drive geostrophic circulations \citep[see also ][]{Busse1968,Busse2010,Sauret2012}.   {However,  numerical calculations of \cite{Tilgner2007b} with the stress-free boundary conditions for non-axisymmetric inertial waves show that there are some correlations between the azimuthal mean flow and the non-linear force in the bulk. }

We define the mean geostrophic flow as $u_g=\langle \overline{u_\phi}\rangle$, where $\langle \cdot \rangle$ represent an average along the direction of rotation axis. {We shall now analyse how $u_g$ varies with the Ekman number.} 
    
\subsubsection{Geostrophic shear near the critical latitude at the CMB}       
\begin{figure}
\begin{center}
\includegraphics[width=0.7\textwidth]{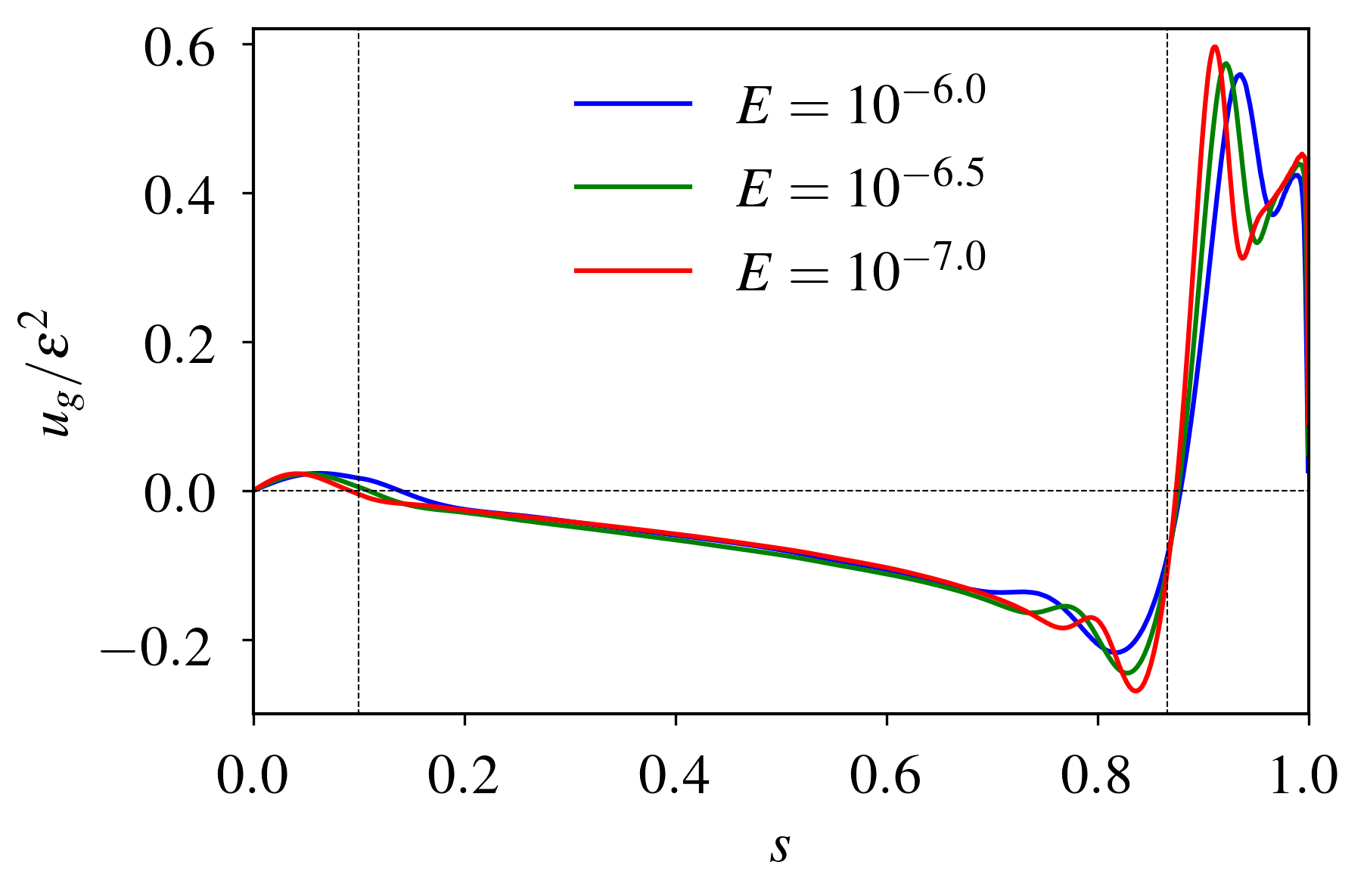} \\
(a)\\
\includegraphics[width=0.7\textwidth]{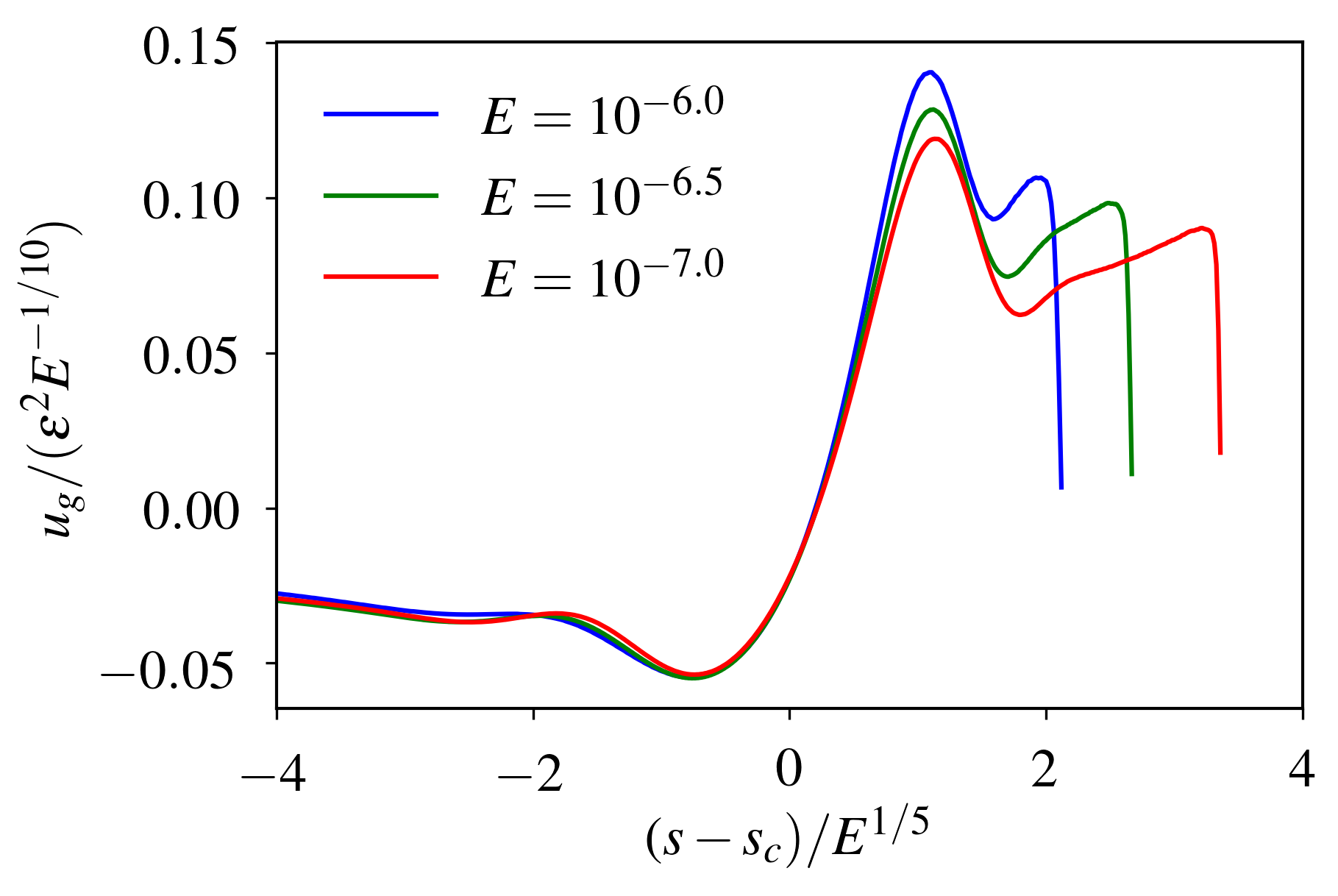}\\
(b)
\caption{(a) Geostrophic velocity profiles at different Ekman numbers in Group~1. Vertical dashed lines represent the tangential cylinder and the cylinder associated with the outer critical latitude. (b) Geostrophic shears rescaled around the cylindrical radius $s_c=0.8660$ associated the outer critical latitude. $u_g$ is rescaled by $\varepsilon^2E^{-1/10}$ and the cylindrical radius is rescale by $E^{1/5}$ (Colour online).}
\label{fig9}
\end{center}
\end{figure}
Theoretical studies \citep{Busse2010, Sauret2012} have shown that the non-linear interactions in the viscous boundary layer of a librating body  can drive a mean zonal flow whose amplitude is independent of the Ekman number and is of order $\mathrm{O}(\varepsilon^2)$ in the absence of inertial waves. When the libration frequency $\omega_L<2$, however, the presence of inertial waves complicates not only the structure of zonal flows but also the dependency on the Ekman number. 

\begin{figure}
\begin{center}
\includegraphics[width=0.95\textwidth]{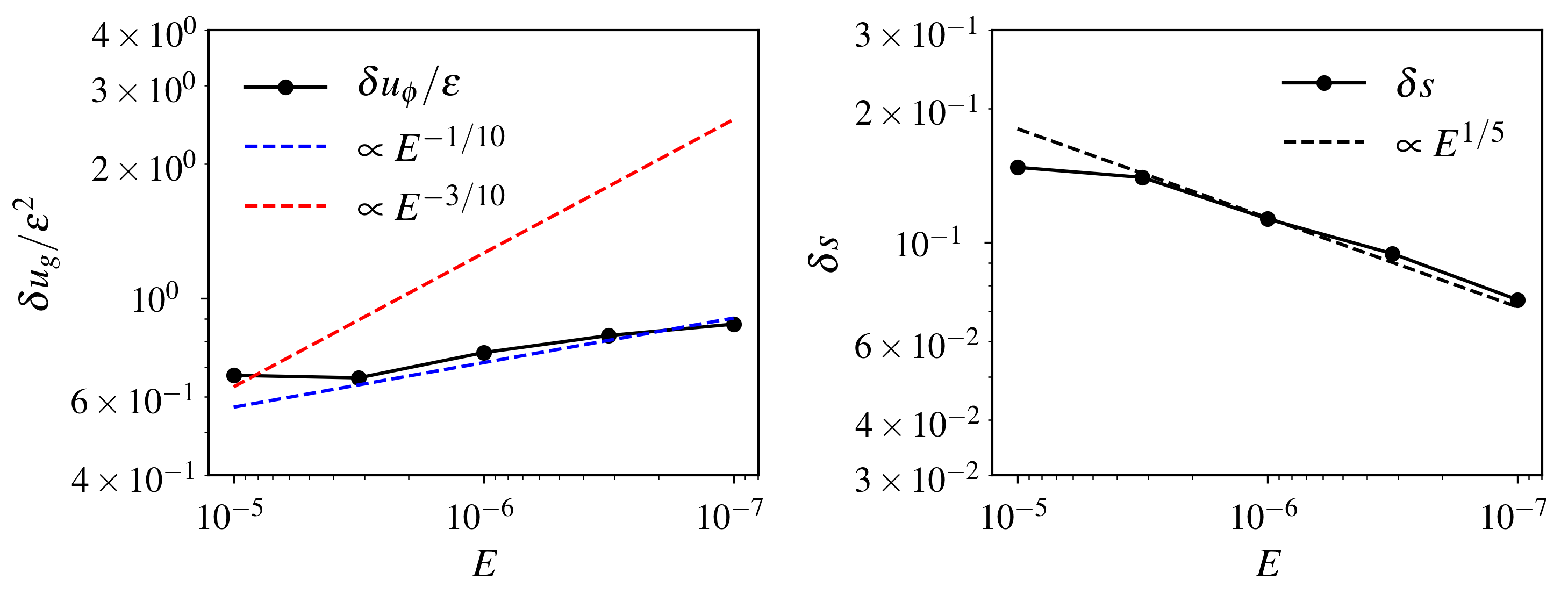} \\
\hspace*{4cm}(a)\hfill (b) \hspace*{4cm}
\caption{(a) Peak-to-peak velocity $\delta u_g$ and (b) distance $\delta s$ between peaks of the geostrophic shear associated with the outer critical latitude as a function the Ekman number in Group~1 (Colour online).}
\label{fig10}
\end{center}
\end{figure}

Figure \ref{fig9}(a) shows geostrophic velocity $u_g$ { as a function of the cylindrical radius $s=r\cos \theta$} at different Ekman numbers in Group~1. For the cylindrical radius $0.2\lesssim s\lesssim 0.7$, the mean zonal flow is independent of the Ekman number. At the cylindrical radius corresponding to the critical latitude, the geostrophic flow increases as the Ekman number reduces while the typical length scale decreases, as observed in precession.  Figure \ref{fig9}(b) shows the rescaled geostrophic  shear associated with the critical latitude. We can see   that the width of the shear scales as $E^{1/5}$ and the velocity amplitude tends to scale as $\varepsilon^2 E^{-1/10}$. {In figure \ref{fig10}, we also plot the peak-to-peak velocity $\delta u_g$ crossing this geostrophic shear and the distance $\delta s$ between the peaks as a function of the Ekman number. We can see that the width of the shear scales as $E^{1/5}$, which is the width of the boundary layer around the critical latitude. Note that at the largest Ekman number in our study ($E=10^{-5}$) the geostrophic shear overlap with the equatorial region exciting a Stewartson Layer that alters the natural scaling at the critical latitude.  The peak-to-peak velocity again shows the scaling $\varepsilon^2 E^{-1/10}$, instead of the scaling $\varepsilon^2 E^{-3/10}$ obtained in a precessing sphere \citep{Noir2001}.} 

At leading order, the flow in { the region around the critical latitude} scales as $u_r\sim \varepsilon E^{1/5} $, $u_\theta \sim \varepsilon $, $u_\phi \sim \varepsilon$, $\upartial/\upartial r \sim E^{-2/5}$ and $\upartial/ \upartial \theta \sim E^{-1/5}$ \citep{Roberts1963,Noir2001}. Hence, the dominant non-linear term in the $\phi$-direction is 
\begin{equation}
u_r\frac{\upartial u_\phi }{\upartial r}\,\sim\, \varepsilon^2 E^{-1/5}\,.
\end{equation}
The integral of the non-linear term $\mathrm{O}(\varepsilon^2 E^{-1/5})$ over the the thickness of  the { region around the critical latitude}  $\mathrm{O}(E^{2/5})$, which gives rise to the non-linear torque $\mathrm{O}(\varepsilon^2 E^{1/5})$.   On the other hand, the viscous torque acting on the ends of geostrophic cylinder scales as $E^{1/2}u_g$. The balance of the {non-linear} and viscous effects leads to a geostrophic flow $u_g$ of $\mathrm{O}(\varepsilon^2 E^{-3/10})$, which is confirmed by numerical calculations in a precessing sphere \citep{Noir2001}. However, this scaling argument {does} not take into account the actual distribution of the integrated quantities.  

\begin{figure}
\begin{center}
\includegraphics[width=0.8\textwidth]{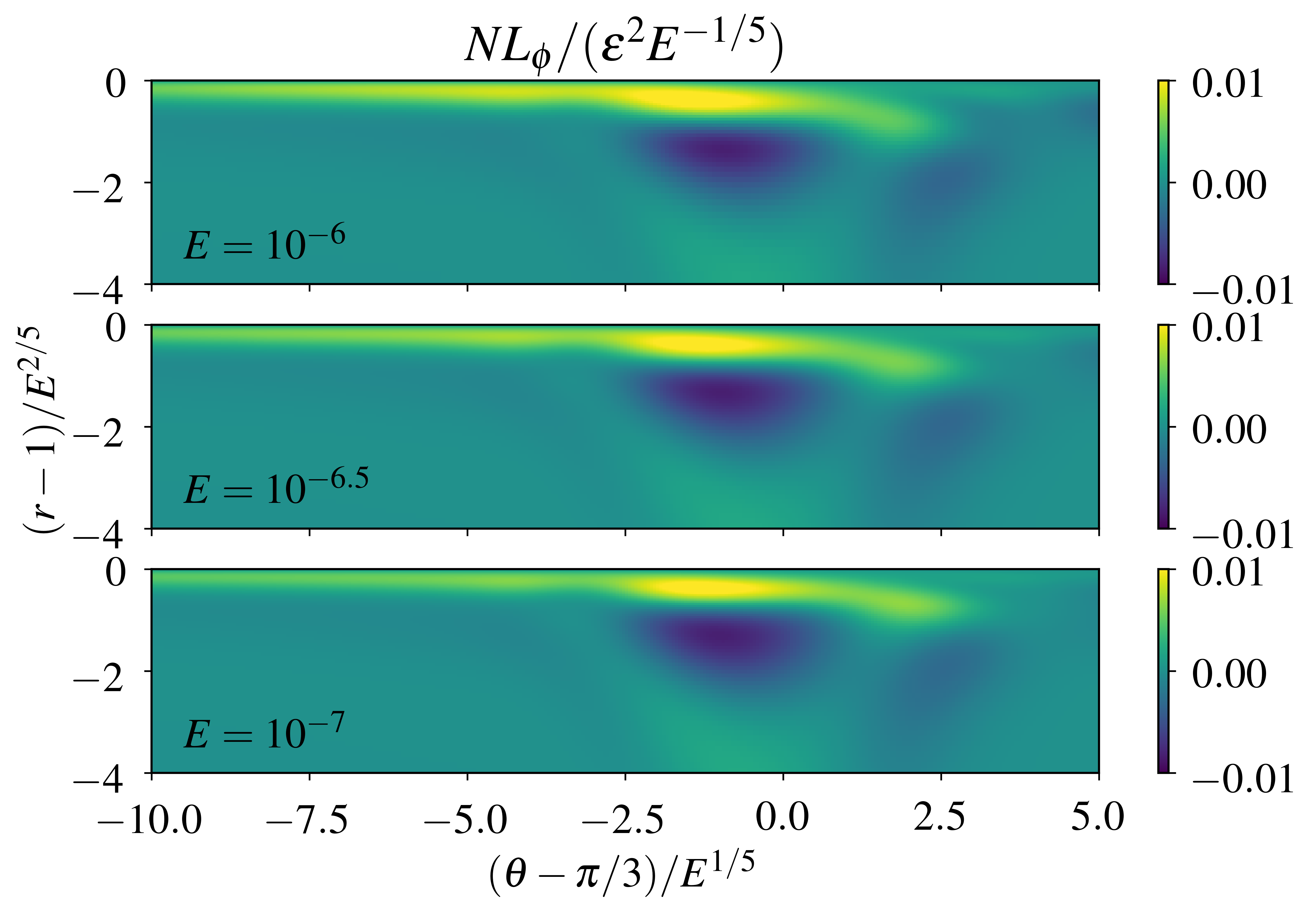}
\caption{Time-averaged non-linear term in the $\phi-$direction $(\overline{\bm{u\cdot \nabla u}})_\phi$, divided by $\varepsilon^2 E^{-1/5}$, in the { region around the critical latitude}  at different Ekman numbers in Group~1. $r$ and $\theta$ are rescaled by $E^{2/5}$ and $E^{1/5}$ respectively (Colour online).}
\label{fig11}
\end{center}
\end{figure}

\begin{figure}
\begin{center}
\includegraphics[width=0.49\textwidth]{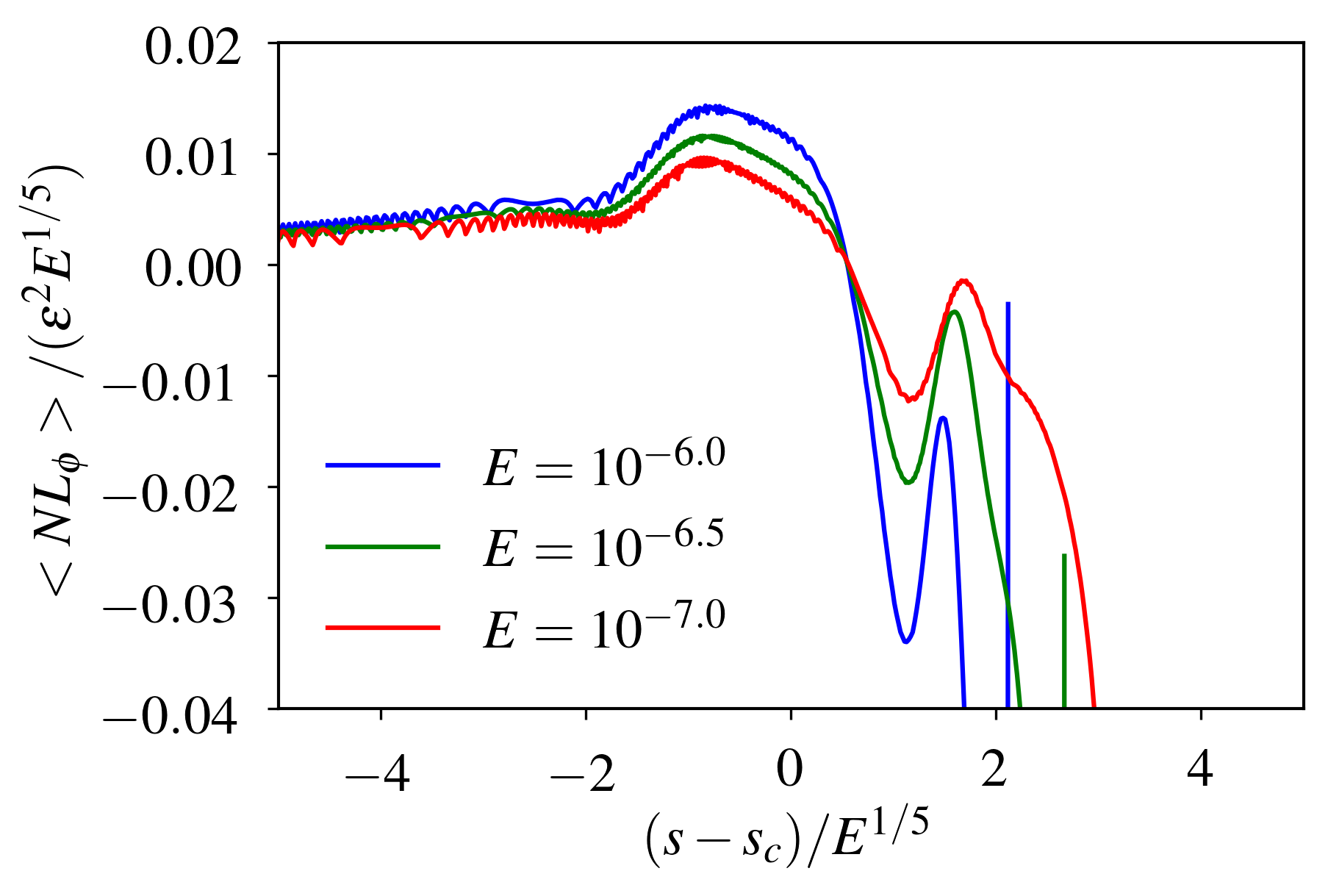}
\includegraphics[width=0.49\textwidth]{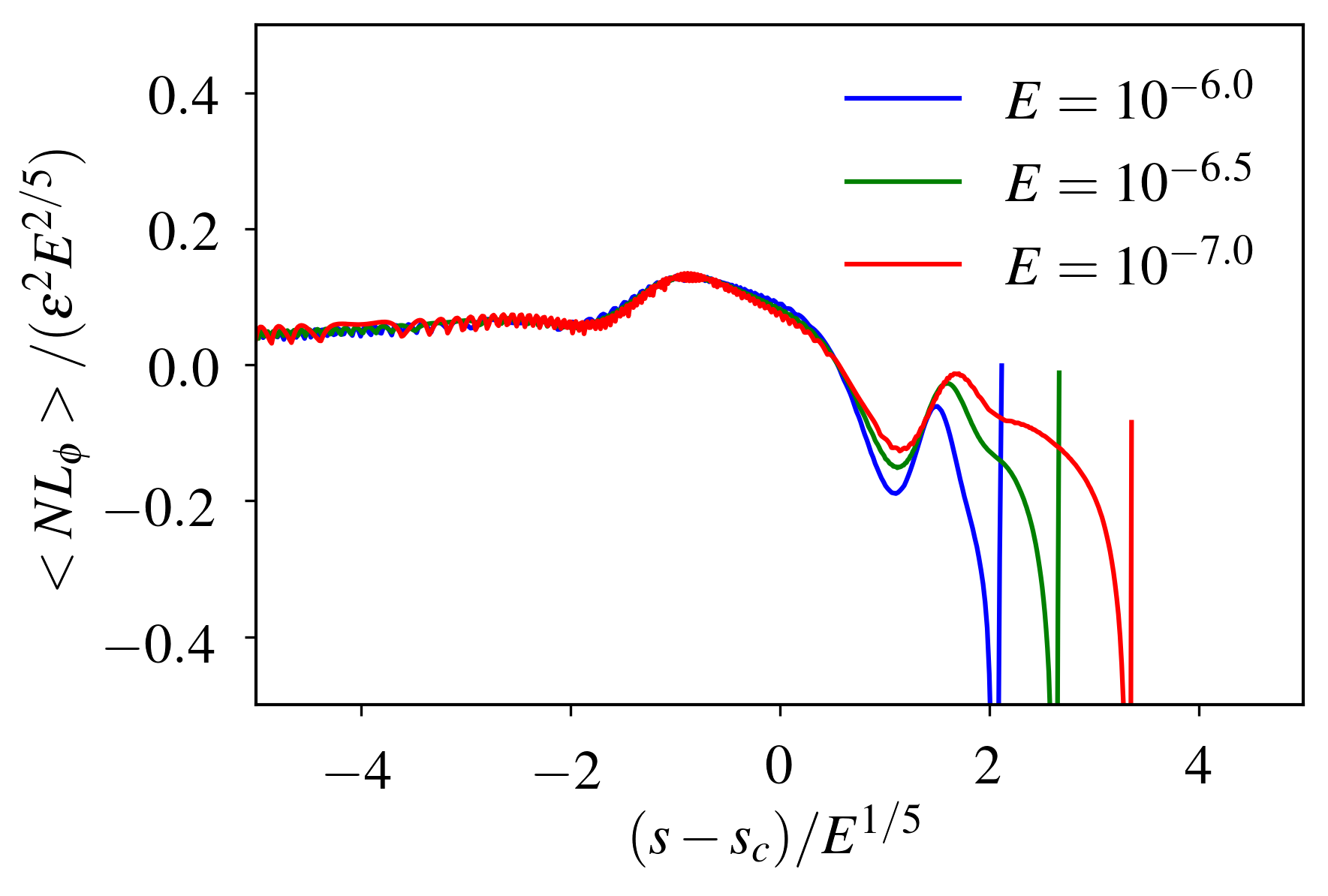}\\
\hspace{4cm}(a)\hfill \hspace{3cm}
\caption{The integral of $(\overline{\bm{u\cdot \nabla u}})_\phi$ along $z$-direction  in Group~1 as a a function of the cylindrical radius $s-s_c$, where $s_c$ is the cylindrical radius of the critical latitude. The integral is rescaled by  $\varepsilon^2E^{1/5}$ in (a) and by  $\varepsilon^2E^{2/5}$ in (b). $s-s_c$ is rescaled by $E^{1/5}$ (Colour online).}
\label{fig12}
\end{center}
\end{figure}

Figure \ref{fig11} shows the time-averaged non-linear term in the $\phi$-direction in the { region around the critical latitude} , rescaled by $\varepsilon^2 E^{-1/5}$. The coordinates $r$ and $\theta$ are rescaled by the by $E^{2/5}$ and $E^{1/5}$ respectively, considering the length scales of the { region around the critical latitude}.  We can see that figure \ref{fig11} confirms the scaling as $\varepsilon^2 E^{-1/5}$ for the non-linear term in the { region around the critical latitude}. Meanwhile, we note that the sign of the non-linear term changes across the boundary layer. Figure \ref{fig12} shows the non-linear term integrated over $z$ as a function of the cylindrical radius  around the geostrophic cylinder at the critical latitude. It turns out that the integrated contribution of the non-linear term scales as $ \varepsilon^2 E^{2/5}$ (at least in the region $s<s_c$; when $s>s_c$, the scaling could be influenced by the the viscous layer at the equator), instead of the predicted scaling $ \varepsilon^2 E^{1/5}$. Our numerical results suggest that the leading order contribution of the non-linear term cancels out when carrying the integral in $z$.  

Indeed, if the leading order non-linear term is cancelled out owing to the integration in the { region around the critical latitude}, the next dominant term in the non-linear term is
\begin{equation}
\frac{u_\theta}{r}\frac{\upartial u_\phi }{\upartial \theta}\,\sim \,\varepsilon^2\,.
\end{equation}
The torque acting on a geostrophic cylinder amounts to $\mathrm{O}(\varepsilon^2 E^{2/5})$,  which is consistent with numerical integral in figure \ref{fig12}(b).  The balance of the second order non-linear torque of order $\mathrm{O}(\varepsilon^2 E^{2/5})$ and the viscous torque of order $\mathrm{O}(E^{1/2}u_g)$ gives rise a geostrophic flow $u_g$ of $\mathrm{O}(\varepsilon^2 E^{-1/10})$, the observed scaling in our numerical calculations.   However, the fundamental reason for the non-linear term canceling out when integrated in $z$ for the libration case but not for the precession remains an open question.

\subsubsection{Geostrophic shear near the critical latitude at the ICB.}

\begin{figure}
\begin{center}
\includegraphics[width=0.7\textwidth]{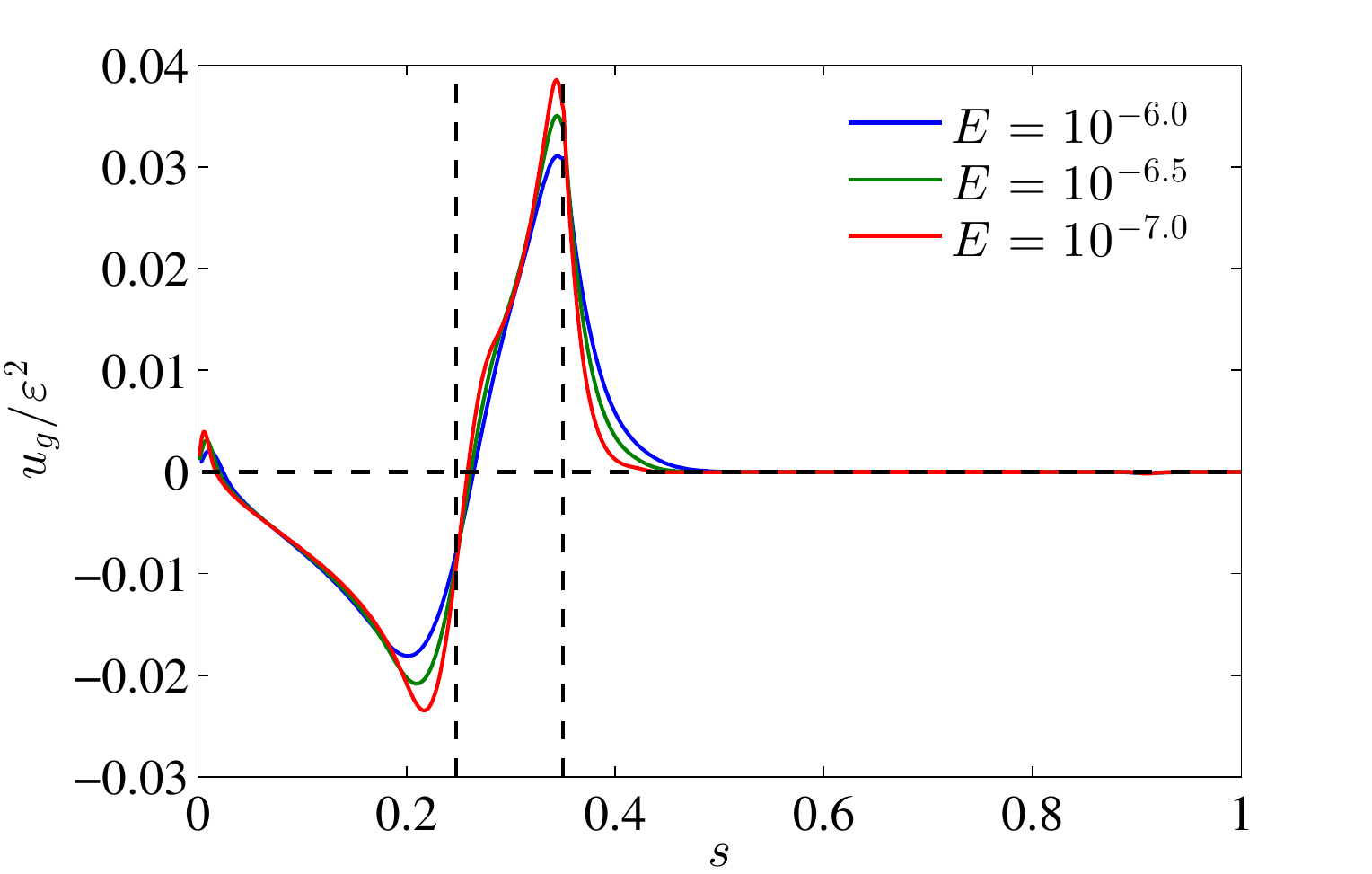} \\
(a) \\
\hspace*{-1.2cm}\includegraphics[width=0.68\textwidth]{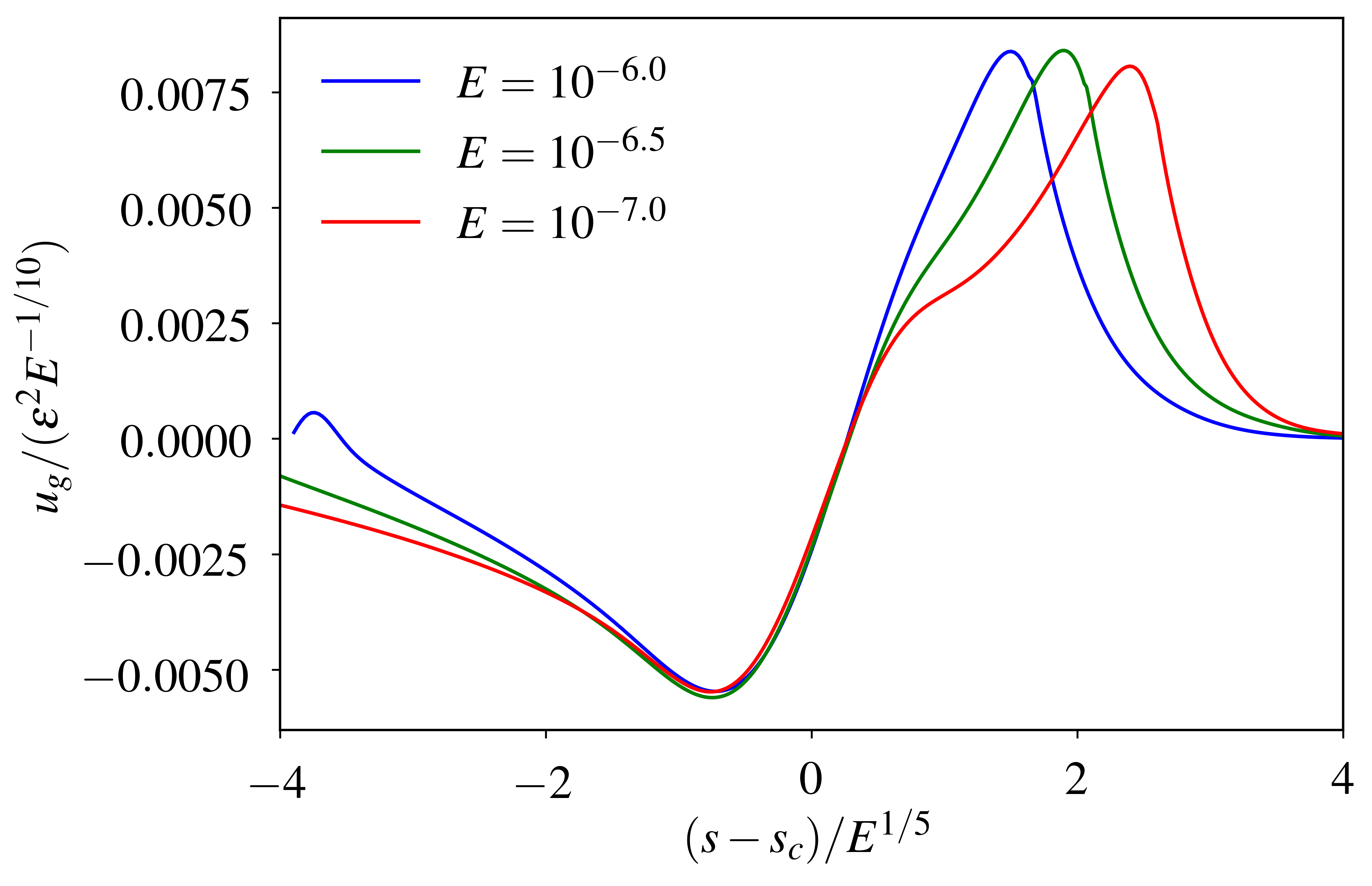} \\
(b) 
\caption{(a) Geostrophic velocity profiles at different Ekman numbers in Group~2. Vertical dashed lines represent the cylinder associated with the inner critical latitude and the tangential cylinder. (b) Geostrophic shears rescaled around the cylindrical radius $s_c=0.3083$ associated with the inner critical latitude. $u_g$ is rescaled by $\varepsilon^2E^{-1/10}$ and the cylindrical radius is rescale by $E^{1/5}$ (Colour online). }
\label{fig13}
\end{center}
\end{figure}

Figure \ref{fig13}(a) shows geostrophic velocity profiles at different Ekman numbers in Group~2 where only the inner core is librating. The geostrophic flow is almost absent outside the tangential cylinder as the outer shell is stationary in the rotating frame. There exists a shear layer attached to the tangential cylinder and another one attached to the inner critical latitude. 

\begin{figure}
\begin{center}
\includegraphics[width=0.8\textwidth]{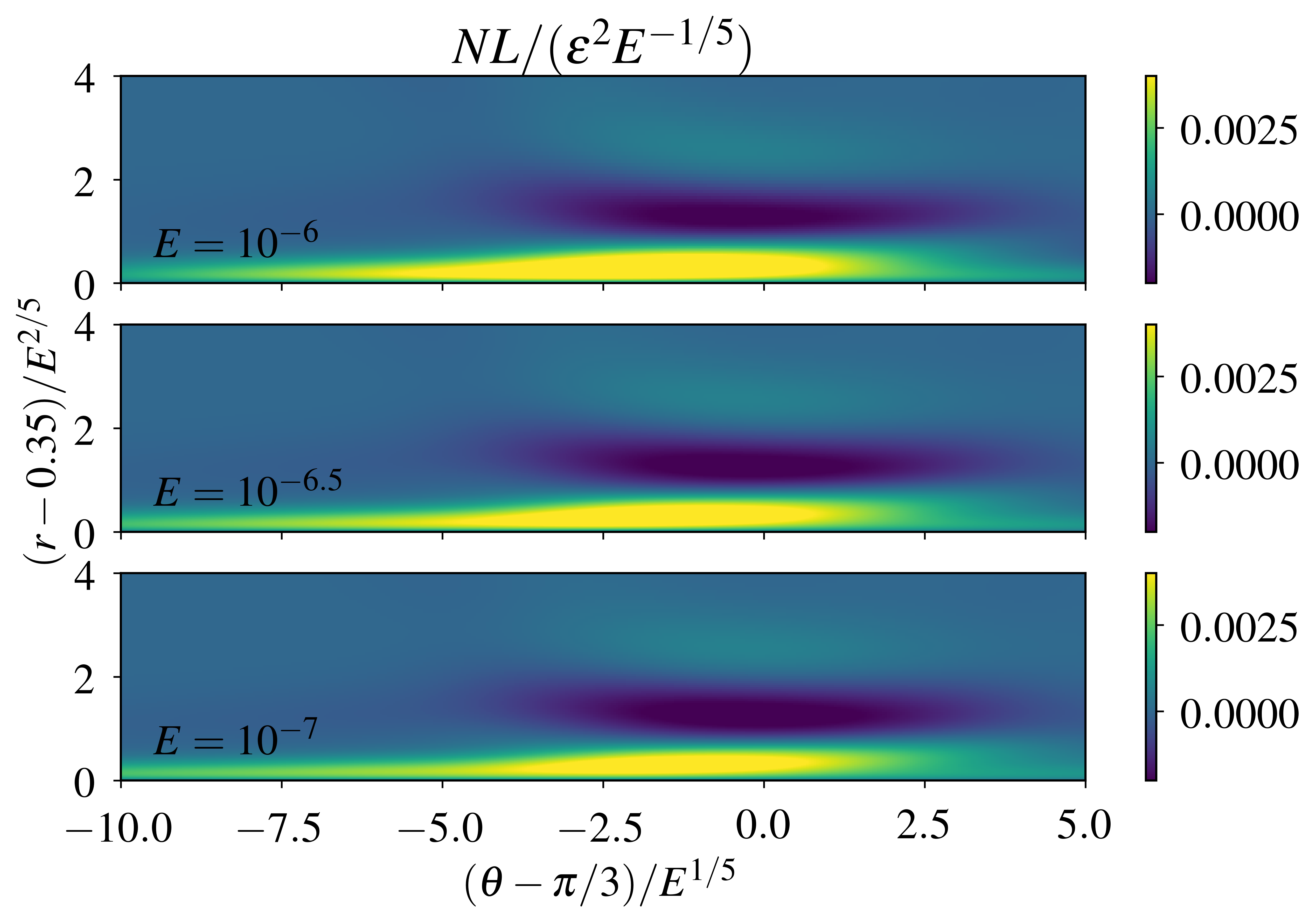}
\caption{ Time-averaged non-linear term in the $\phi-$direction $(\overline{\bm{u\cdot \nabla u}})_\phi$, divided by $\varepsilon^2 E^{-1/5}$, in the { region around the critical latitude}  at different Ekman numbers in Group~2. $r$ and $\theta$ are rescaled by $E^{2/5}$ and $E^{1/5}$ respectively (Colour online).}
\label{fig14}
\end{center}
\end{figure}

\begin{figure}
\begin{center}
\includegraphics[width=0.7\textwidth]{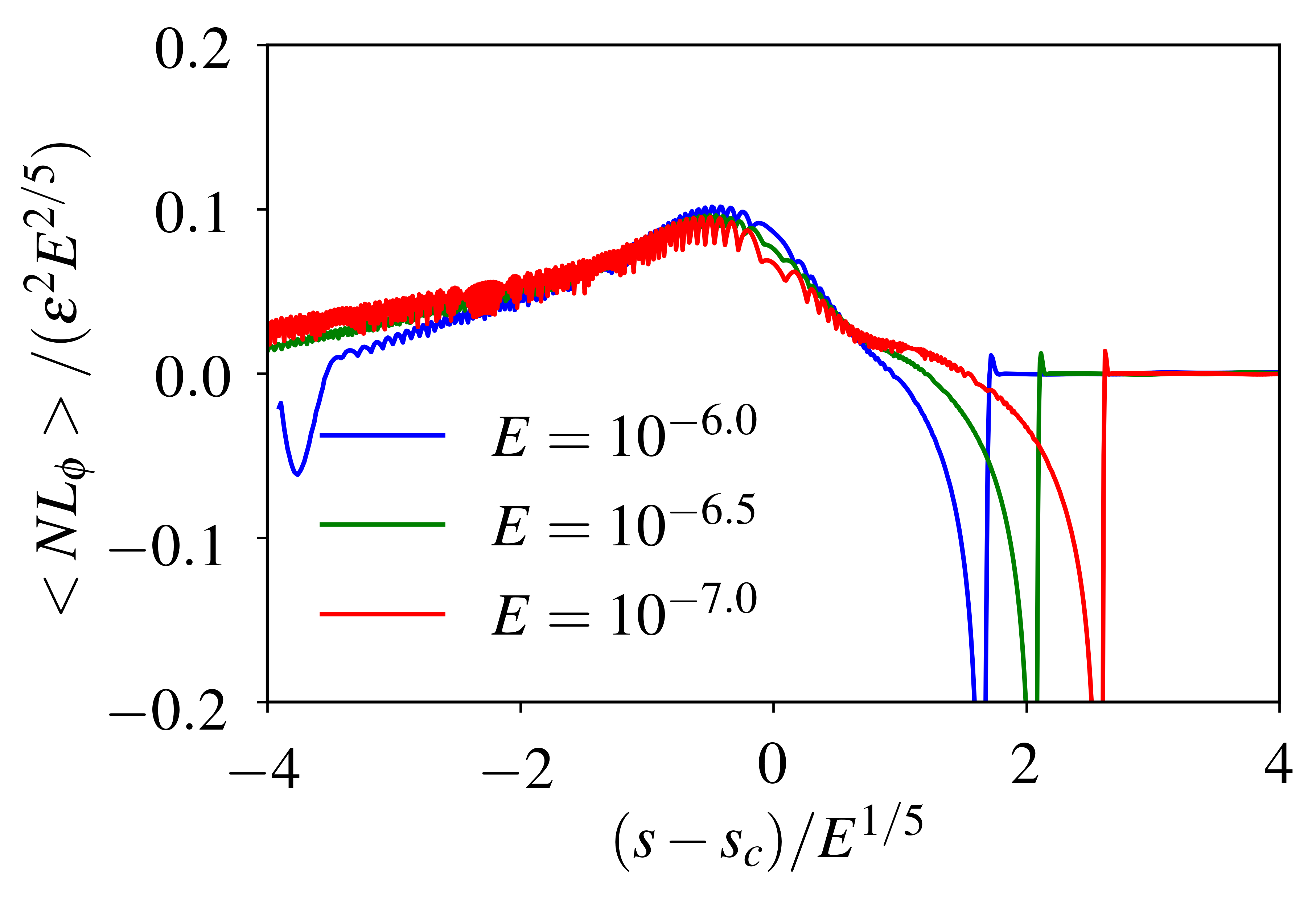}
\caption{The integral of $(\overline{\bm{u\cdot \nabla u}})_\phi$ along $z$-direction in Group~2 as a a function of the cylindrical radius $s-s_c$, where $s_c$ is location of the critical latitude. The integral and $s-s_c$ are rescaled by $\varepsilon^2E^{2/5}$ and $E^{1/5}$ respectively (Colour online) .}
\label{fig15}
\end{center}
\end{figure}

At the critical latitude at the ICB, the flow in the boundary layer {has} the same scaling as at CMB \citep{LeDizes2017}. While the emission of the tangential inertial waves is thinner and more intense in comparison with the radially reflected wave we expect the scaling for the non-linear interaction to be identical to those obtained at the CMB, i.e. $(\overline{\bm{u\cdot \nabla u}})_\phi$ is  $\mathrm{O}(\varepsilon^2 E^{-1/5})$ and $\mathrm{O}(\varepsilon^2)$ at leading and next order, respectively. We confirm these scalings in figure \ref{fig14} that shows the time-averaged non-linear term in the $\phi$-direction in the { region around the critical latitude}  at the inner boundary in Group~2. Similarly to what is observed in Group~1, the spatial distribution of the non-linear term leads to a torque acting on a geostrophic cylinder that vanishes at leading order while the next order contribution amounts to $\mathrm{O}(\varepsilon^2 E^{2/5})$ as shown in  figure \ref{fig15}. Following the same analysis as for the geotrophic shear in Group~1 at the CMB we thus expect the geostrophic shear to be $\mathrm{O}(\varepsilon^2E^{-1/10})$ in amplitude over a width of  $\mathrm{O}(E^{1/5})$. In figure \ref{fig13}(b) representing the rescaled geostrophic flow near the critical latitude, { we} see that the curves at different Ekman numbers are well collapsed, confirming the proposed scaling. We note that the shear layer attached to the tangential cylinder overlap with the one produce at the critical latitude. {Calculation at} much lower Ekman are necessary to disentangle these two.

\subsubsection{Geostrophic shear from inertial waves reflections.}
\begin{figure}
\begin{center}
\includegraphics[width=0.6\textwidth]{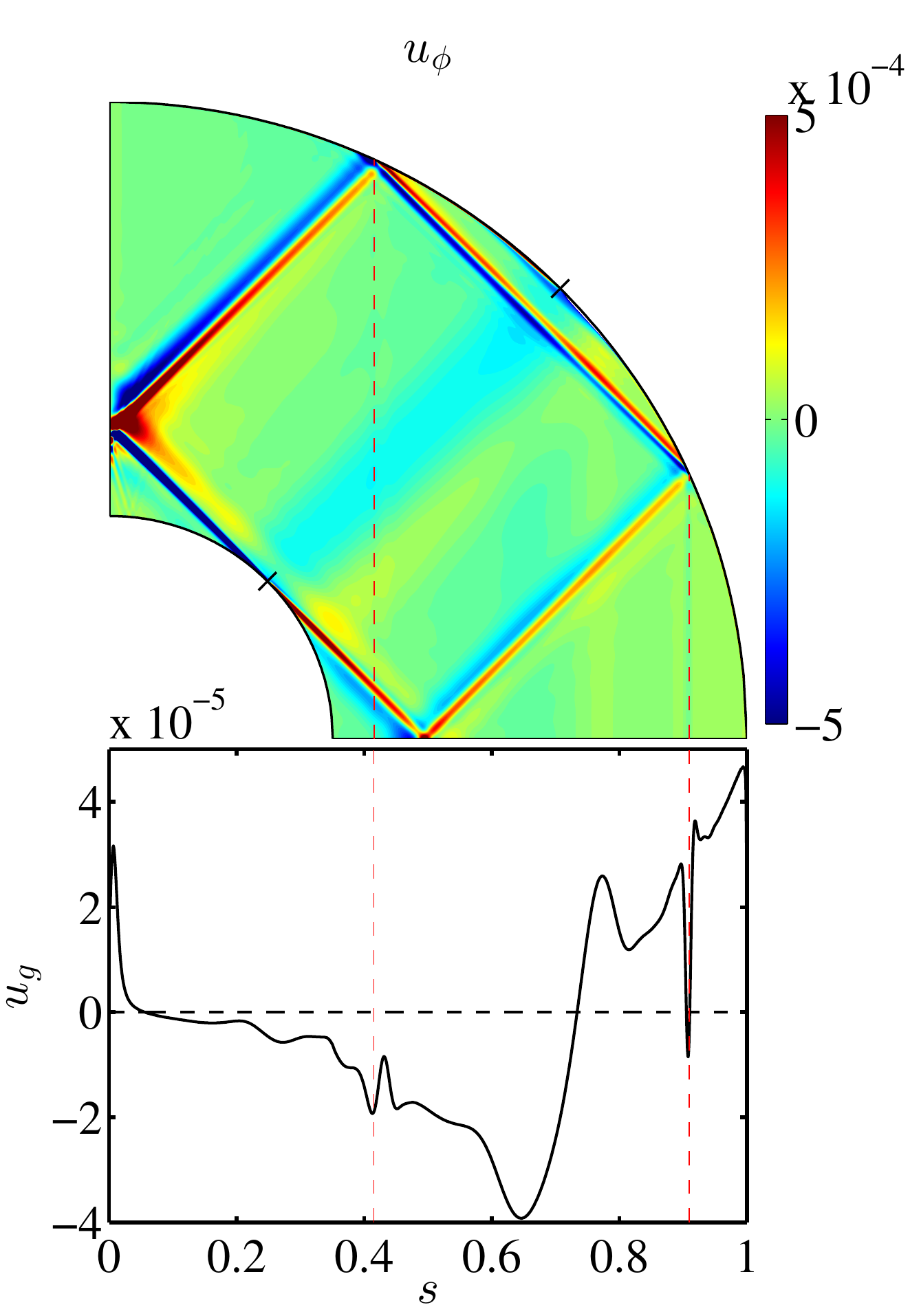}
\caption[Azimuthal velocities $u_\phi$ (top)  in the meridional plane at $E=10^{-7}$ in Group~3 and the corresponding geostrophic velocity profile (bottom).]{Azimuthal velocities $u_\phi$ (top)  in the meridional plane at $E=10^{-7}$ in Group~3 and the corresponding geostrophic velocity profile (bottom). Vertical red dashed lines indicate cylindrical radius of the reflection positions of the conical shear shear layers (Colour online). }
\label{fig16}
\end{center}
\end{figure}

Figure \ref{fig16} shows the geostrophic velocity $u_g$ as a function of cylindrical radius $s=r \cos \theta$ for a case in Group~3, where the inner core is uniformly rotating and the outer shell is librating at $\omega_L=1.4242$. While the geostrophic shear around $s=0.7071$  is associated with the critical latitude, from which inertial waves are initially launched, two additional geostrophic shears around $s=0.4149$ and $s=0.9099$ (two red dashed lines) clearly correspond to the locations where inertial waves are reflected on the outer boundary. Note that inertial waves are also reflected on the outer boundary in the cases of Group~2 where only the inner core is librating. However, there is no visible geostrophic shears associated with the reflection of inertial waves for this case (figure \ref{fig8}(b)), for which the outer shell is stationary in the rotating frame, i.e. there is no viscous layer at the outer boundary at leading order. This suggests that the mean zonal flows arise from the {non-linear} interactions of the viscous flows driven by both the reflecting inertial waves and the background libration. In figure \ref{fig16}, we note also a strong jet near the rotation axis, which results from two crossing wave beams on the rotation axis as pointed out by \cite{LeDizes2017}.

\begin{figure}
\begin{center}
\includegraphics[width=0.7\textwidth]{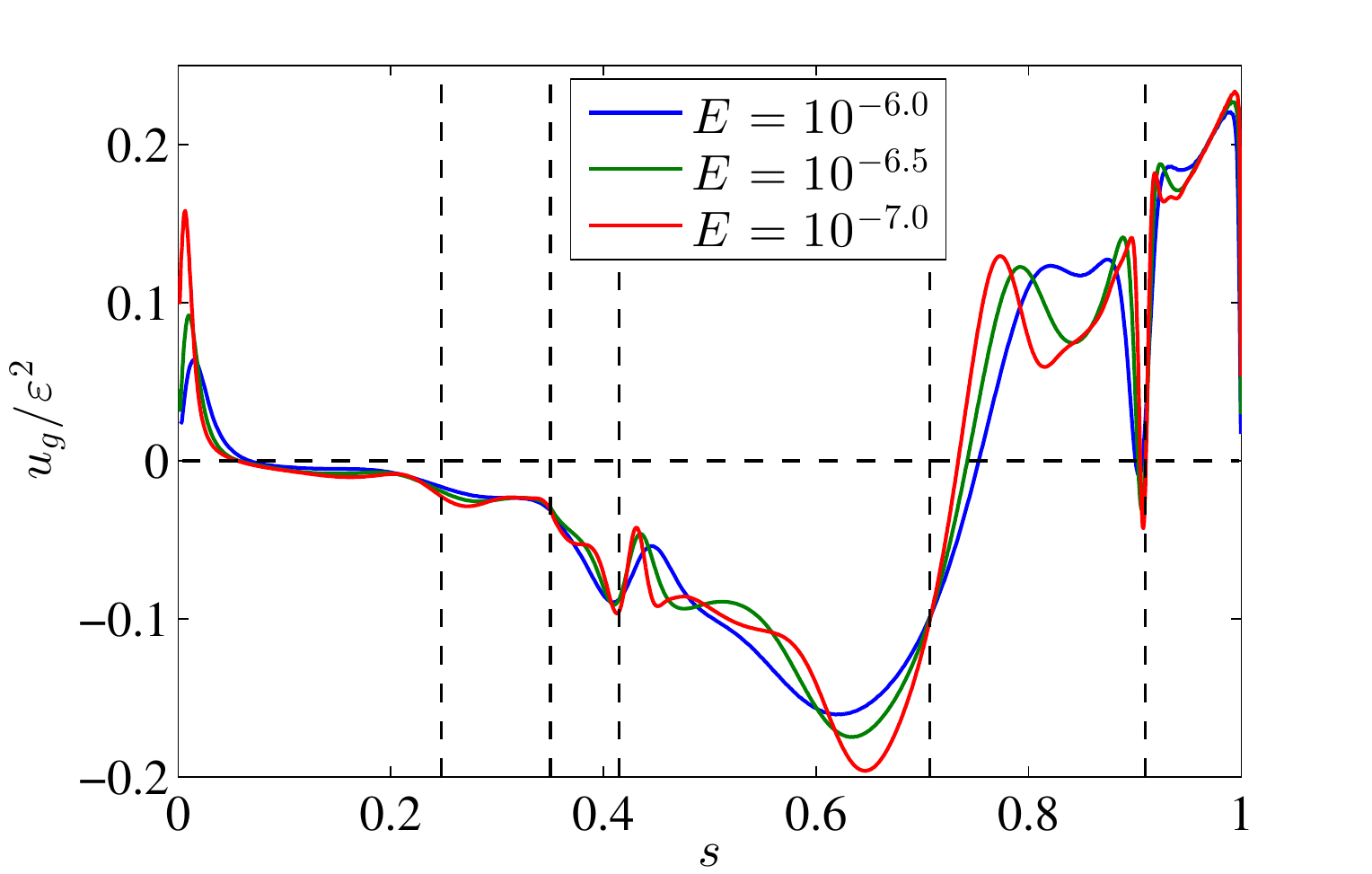}\\
(a)\\
\hspace*{-1cm}\includegraphics[width=0.68\textwidth]{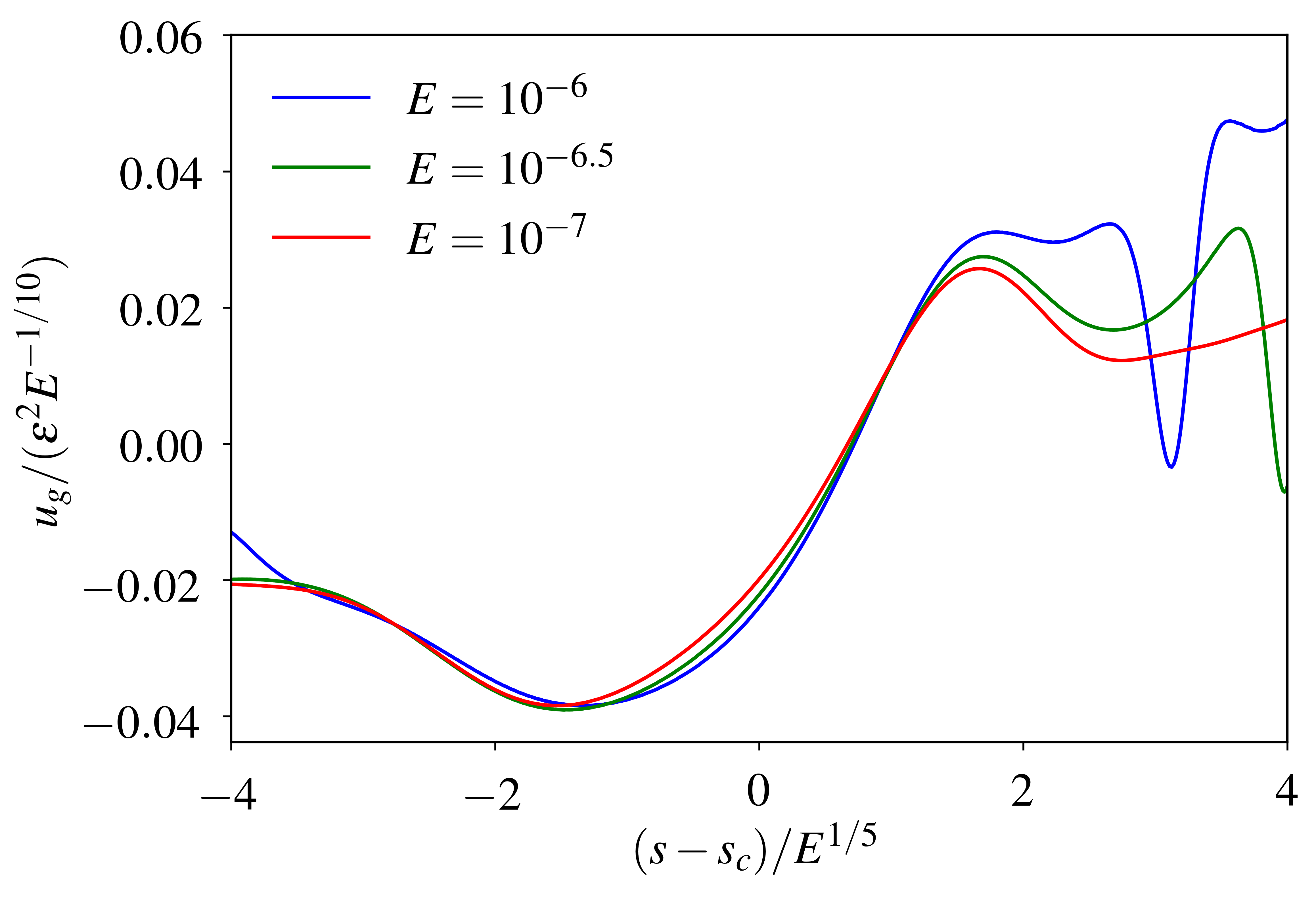} \\
(b) 
\caption{ (a) Geostrophic velocity profiles at different Ekman numbers in Group~3. Vertical dashed lines (from left to right) represent the inner critical latitude, the tangential cylinder, the reflection position close to the rotation axis, the outer critical latitude and the reflection position close to the equator. (b) Geostrophic shear rescaled around the cylindrical radius $s_c=0.8660$ associated the outer critical latitude. $u_g$ is rescaled by $\varepsilon^2E^{-1/10}$ and the cylindrical radius is {rescaled} by $E^{1/5}$ (Colour online). }
\label{fig17}
\end{center}
\end{figure}

Figure \ref{fig17}(a) shows geostrophic velocity profiles at different Ekman numbers in Group~3, significantly more complex than previous two cases. Several geostrophic shears are generated owing to the critical latitude, the reflections of inertial waves and the crossing of two wave beams on the rotation axis. Figure \ref{fig17}(b) focuses on the geostrophic shear near the outer critical latitude, which as expected exhibits a scaling $\varepsilon^2E^{-1/10}$ in  amplitude and  $E^{1/5}$  in cylindrical radius. 

\begin{figure}
\begin{center}
\includegraphics[width=0.95\textwidth]{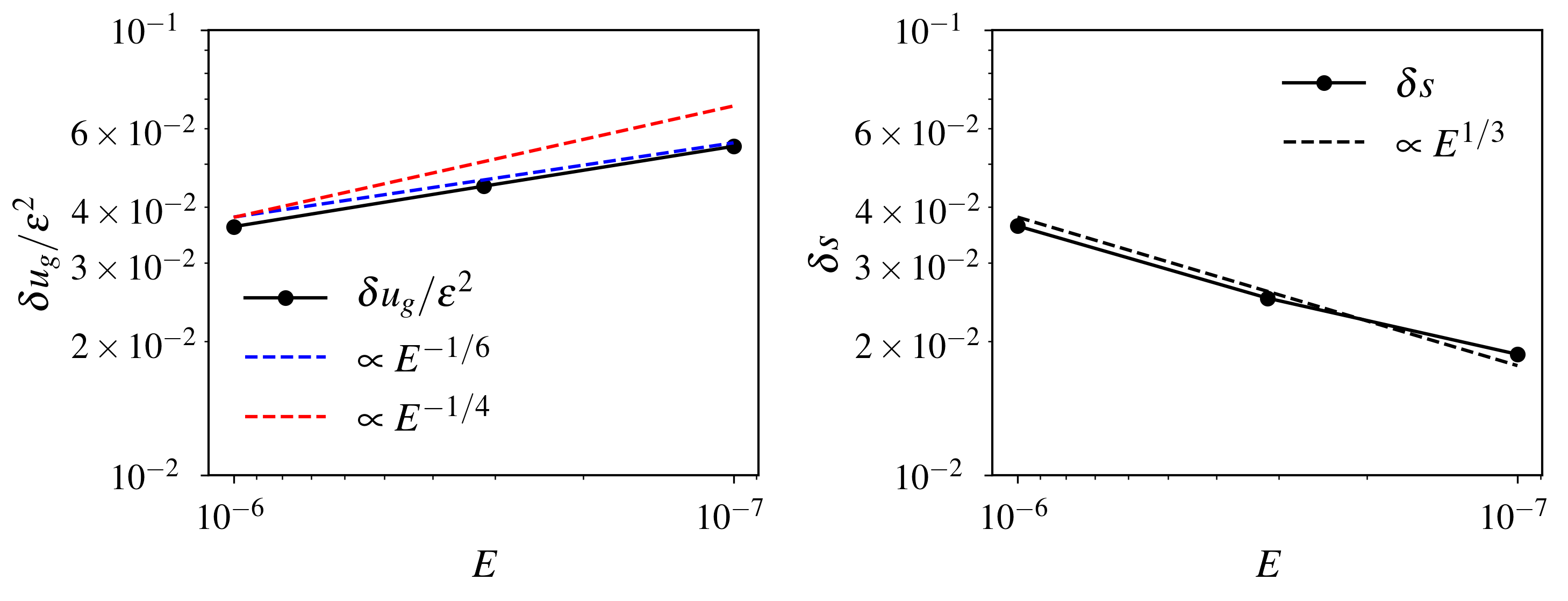} \\

\hspace*{4cm}(a)\hfill (b) \hspace*{3cm}
\caption{(a) Peak-to-peak velocities $\delta u_g$ and (b) distances $\delta s$ between peaks of the geostrophic shear associated with the wave reflection (close to the rotation axis) as a function the Ekman number in Group~3 (Colour online). }
\label{fig18}
\end{center}
\end{figure}

At a cylindrical radius $s=0.4149$, the geostrophic shear is due to the reflection of inertial waves. We extract the peak-to-peak velocities $\delta u_g$ and the distances between the peaks in figure  \ref{fig18}. We can see that the width of the reflection induced geostrophic shears scales as $E^{1/3}$, which is the same as the width of the reflected inertial waves, while the amplitude scales as $\varepsilon^2 E^{-1/6}$ (figure \ref{fig18}(a)).

{We can apply a similar scaling analysis as for the critical latitude to derive the amplitude and width of the geostrophic shear due to the reflection. However, we cannot yet draw a hard conclusion on the scaling of the incoming waves spawned from the critical latitude at the ICB. We may consider the two scalings of the velocity amplitude  $\varepsilon E^{1/6}$ and $\varepsilon E^{1/12}$ for the incoming inertial waves. Using the first scaling leads to an amplitude of the geostrophic velocity $\mathrm{O}(\varepsilon^2 E^{-1/3})$ at leading order and $\mathrm{O}(\varepsilon^2E^{-1/6})$ at the next order. Using the second amplitude leads to a geostrophic velocity amplitude $\mathrm{O}(\varepsilon^2 E^{-5/12})$ at leading order and $\mathrm{O}(\varepsilon^2E^{-1/4})$ at the next order. While our numerical results favor a scaling with $\varepsilon E^{1/12}$ for the incoming waves, the observed geostrophic velocities suggest a mechanism based on a second order non-linear interaction with waves of amplitudes $\mathrm{O}(\varepsilon E^{1/6})$, assuming a similar spatial cancellation occurs as for the critical latitude at leading order. This contradiction clearly shows that further investigations at much lower Ekman numbers are necessary to conclude. The width of the geostrophic shear scales as $E^{1/3}$ (figure \ref{fig18}(b)), which is the length scale of the inertial waves being reflected.}

\section{Discussion} \label{sec3:diss}

Motivated by understanding the dynamics in planetary cores and subsurface oceans, we numerically investigated fluid flows in longitudinally librating spherical shells. 
We investigate the Ekman number dependencies of the oscillating conical shear layers and the steady geostrophic shears. We observe geostrophic shears resulting from non-linear interactions in the boundary layers near the critical latitudes at the ICB and at the CMB. In addition we report the reflections of inertial waves in the viscous boundary layers can {drive} significant geostrophic shears. 
    
The time-dependent flows in the bulk mainly consist of conical shear layers spawned from the critical latitudes. For the libration frequencies $\omega_L=1.0$ and $\omega_L=\sqrt{2}$, the conical shears form simple trajectories, which allow us to investigate their structure in detail. Our numerical results show that the conical shear layers spawned from the outer critical latitudes carry velocities  $\mathrm{O}(\varepsilon E^{1/5})$ with a width  $\mathrm{O}(E^{1/5})$, in agreement with previous theoretical predictions \citep{Noir2001,Kida2011}. 
{  For the conical shear shear layers spawned from the inner critical latitudes, we confirm that the width scales as  $\mathrm{O}(E^{1/3})$ as predicted \citep{Kerswell1995}.  Regarding the velocity amplitude, our numerical results favor the scaling $\mathrm{O}(\varepsilon E^{1/12})$ predicted by \cite{LeDizes2017} over the scaling $\mathrm{O}(\varepsilon E^{1/6})$ initially suggested by \cite{Kerswell1995}, though the Ekman numbers in our calculations are not asymptotically small enough to pin down the amplitude scaling. Note that previous theoretical predictions are based on an unbounded fluid domain. It remains unclear whether these predicted scalings can  be applied to closed containers without adjustments.}
  
The non-linear interactions of the time-dependent flows in the viscous boundary layers can drive a steady mean zonal flow in the bulk. In the absence of inertial waves, the magnitude of the mean zonal flow is proportional to the square of the libration amplitude and is independent of the Ekman number \citep{Busse2010,Sauret2012}. 
However, the excitation of inertial waves leads to more complicated zonal flows in terms of both their amplitude and structure. In particular, we observed several different geostrophic shear layers associated with inertial waves. {We show that the geostrophic shears near the critical latitudes at the ICB and at the CMB exhibiting the same scalings,  i.e. the width of $ O(E^{1/5})$ and the amplitude of $\mathrm{O}(\varepsilon^2 E^{-1/10})$, despite different conical shear layers are spawned from the critical latitudes at ICB and at the CMB. } However, the velocity amplitude of the geostrophic shear contrasts with $\mathrm{O}(\varepsilon^2 E^{-3/10})$ observed in precessing sphere \citep{Noir2001}. We show that this difference is probably due to the spatial distribution of the non-linear interactions in the boundary layer, that cancels out when integrating over a geostrophic cylinder for libration. 

Apart from the geostrophic shear associated with the critical latitudes, we found that the reflections of inertial waves in the viscous boundary layers can drive a geostrophic shear. The width of the geostrophic shear due to the reflection scales as $\mathrm{O}(E^{1/3})$, which is also the length scale of the inertial waves being reflected. The velocity amplitude tends to scale as $\mathrm{O}(\varepsilon^2 E^{-1/6})$ based on our numerical results, yet theoretical analysis remain to be elucidated. {Nevertheless, it is evident that the amplitude of the geostrophic shear increases as reducing the Ekman number. Simply applying these scaling to planetary cores or subsurface oceans, we estimate that typical zonal velocities of $10^{-5}\sim 10^{-3}$ m s$^{-1}$ can be generated due to the reflection of libration-driven inertial waves based on planetary parameters given in \cite{Noir2009}. For comparison, the typical velocity in the Earth's outer core is about $10^{-4}$ m s$^{-1}$ inferred from the geomagnetic secular variations \citep[e.g.][]{Barrois2018}. }

{ In this paper, the libration frequencies have been chosen such that the conical shear layers form simple beam structures (periodic orbits). However, we note that fluid responses depend on the libration frequency. In particular, wave attractors exist in spherical shells in certain frequency bands, which we did not consider in this study.  \cite{Rieutord2018} have shown that multiple different length scales exist within the shear layers associated with inertial wave attractors. The amplitude of the forced wave attractors remains to be investigated.  Steady mean flows can be generated by the non-linear interactions of wave attractors \citep{Maas2001}, though they may have a different nature.}   
   
{The libration amplitude is set to be small such that no fluid instabilities are triggered in our numerical calculations. This allows us to  perform axisymmetric simulations} and  characterize several shear layers in the bulk of fluid and to investigate the Ekman number dependencies of these shear layers. However, these shear layers may become unstable, leading to more complicated flows and even turbulence. \cite{Noir2009} have shown that turbulent flows develop in the vicinity of boundary owing to the centrifugal instabilities in a librating spherical shell. Extrapolation of their experimental results suggests that libration-driven turbulence is expected in some librating planetary bodies \citep{Noir2009}. Apart from the boundary layer instabilities, \cite{Lin2015} have shown that the conical shear layers in a precessing sphere can become unstable through parametric instabilities, similar instabilities should be expected in librating bodies as well. Furthermore, { the geostrophic shear layers can become unstable at small Ekman numbers due to a shear instability \citep{Sauret2014}.}  As the amplitude of all geostrophic shears {increases} with decreasing the Ekman number, the geostrophic shears are very likely to become unstable in planetary liquid cores or subsurface oceans, contributing to the energy dissipation and and exchanges of the angular momentum in these {systems}. 

\section*{Acknowledgements}
 We would like to thank two anonymous referees for their constructive comments that helped to improve the paper. The numerical code used in this study is provided by M. Calkins. Some of numerical calculations were carried out on the Brutus cluster at ETH Zurich and on clusters at Swiss National Supercomputing Center (CSCS) under the Project No. s872 . Y.L. is supported by the B-type Strategic Priority Program of the Chinese Academy of Sciences (XDB41000000), the pre-research project on Civil Aerospace Technologies ofChina National Space Administration (No. D020308) and by the National Natural Science Foundation ofChina (grant No. 41904066). J.N was partially funded through SNF grants \#20021-140708 and \#20021-165641. This study 
has received funding from the European Research Council 
(ERC) through  Grant No. 247303 (MFECE)
under the European Union’s Horizon 2020 research and innovation programme
at ETH Zurich.



\end{document}